\documentclass[preprint,
superscriptaddress,
showpacs,
amsmath,amssymb,
aps,
prc,
]{revtex4-1}
\usepackage{graphicx}
\usepackage{color}
\usepackage{epstopdf}
\usepackage{makecell,multirow,diagbox}
\usepackage{setspace}

\begin{document}

\title{Feasibility of studying astrophysically important charged-particle emission with
the Variable Energy $\gamma$-ray system (VEGA) at Extreme Light Infrastructure -- Nuclear Physics (ELI-NP)}

\author{H.Y. Lan}
 \affiliation{School of Nuclear Science and Technology, University of South China, Hengyang, 421001, China}
\author{W. Luo}%
 \email{wen.luo@usc.edu.cn}
 \affiliation{School of Nuclear Science and Technology, University of South China, Hengyang, 421001, China}
 \author{Y. Xu}%
 \email{yi.xu@eli-np.ro}
 \affiliation{Extreme Light Infrastructure - Nuclear Physics (ELI-NP), Horia Hulubei National Institute for R\&D in Physics and
Nuclear Engineering (IFIN-HH), 30 Reactorului Str., 077125 Buchurest-Magurele, Romania}
 \author{D.L. Balabanski}%
 \affiliation{Extreme Light Infrastructure - Nuclear Physics (ELI-NP), Horia Hulubei National Institute for R\&D in Physics and
Nuclear Engineering (IFIN-HH), 30 Reactorului Str., 077125 Buchurest-Magurele, Romania}
\author{G.L. Guardo}
 \affiliation{INFN-Laboratori Nazionali del Sud, 95123, Catania, Italy}
\author{M. La Cognata}
 \affiliation{INFN-Laboratori Nazionali del Sud, 95123, Catania, Italy}
\author{D. Lattuada}
 \affiliation{INFN-Laboratori Nazionali del Sud, 95123, Catania, Italy}
 \affiliation{Universit\`{a} degli Studi di Enna KORE Viale delle Olimpiadi, 94100, Enna, Italy}
\author{C. Matei}
 \affiliation{Extreme Light Infrastructure - Nuclear Physics (ELI-NP), Horia Hulubei National Institute for R\&D in Physics and
Nuclear Engineering (IFIN-HH), 30 Reactorului Str., 077125 Buchurest-Magurele, Romania}
 \author{R.G. Pizzone}
 \affiliation{INFN-Laboratori Nazionali del Sud, 95123, Catania, Italy}
\author{T. Rauscher}
\affiliation{Department of Physics, University of Basel, 4056 Basel, Switzerland}
\affiliation{Centre for Astrophysics Research, University of Hertfordshire, Hatfield AL10 9AB, United Kingdom}
\author{J.L. Zhou}%
 \affiliation{School of Nuclear Science and Technology, University of South China, Hengyang, 421001, China}


\begin{abstract}
In the environment of a hot plasma, as achieved in stellar explosions, capture and photodisintegration reactions proceeding on excited states in the nucleus can considerably contribute to the astrophysical reaction rate. Usually, such reaction rates including the excited-state contribution are obtained from theoretical calculations as the direct experimental determination of these astrophysical rates is currently unfeasible. Future experiments could provide constraining information on the current reaction models which would improve the predictive power of the theoretical reaction rates. In the present study, experiments of photodisintegration with charged-particle emission leading to specific excited states in the residual nucleus are proposed. The expected experimental results can be used to determine the particle-transmission coefficients in the model calculations of photodisintegration and capture reactions. With such constrained transmission coefficients, the astrophysical reaction rates especially involving the excited state contributions can be better predicted and implemented in astrophysical simulations.
In particular, ($\gamma$,p) and ($\gamma$,$\alpha$) reactions in the mass and energy range relevant to the astrophysical $p$ process are considered and the feasibility of measuring them with the ELISSA detector system at the future Variable Energy $\gamma$-ray (VEGA) facility at ELI-NP (Extreme Light Infrastructure - Nuclear Physics) is investigated. To this end, in a first step 17 reactions with proton emission and 17 reactions with $\alpha$ emission are selected and the dependence of calculated partial cross sections on the variation of nuclear property input is tested.
The simulation results reveal that, for the ($\gamma$,p) reaction on twelve targets of $^{29}$Si, $^{56}$Fe, $^{74}$Se, $^{84}$Sr, $^{91}$Zr, $^{96,98}$Ru, $^{102}$Pd, $^{106}$Cd, and $^{115, 117, 119}$Sn, and the ($\gamma$,$\alpha$) reaction on five targets of $^{50}$V, $^{87}$Sr, $^{123,125}$Te, and $^{149}$Sm, the yields of the reaction channels with the transitions to the excited states in the residual nucleus, namely ($\gamma$,$X_{i}$) with $i\neq$0, are relevant and even dominant. Therefore, these seventeen reactions are considered in the further feasibility study.
For each of the seventeen photon-induced reactions, in order to attain the detectable limit of 100 counts per day for the total proton or $\alpha$-particle yields, the minimum required $\gamma$-beam energies $E_\mathrm{low}$ for the measurements are estimated.
It is further found that for each considered reaction, the total yields of the charged-particle $X$ may be dominantly contributed from one, two or three ($\gamma$,$X_{i}$) channels within a specific, narrow energy range of the incident $\gamma$-beam. If the actual measurements of these photon-induced reactions are performed in this energy range, the sum of the yields of the dominant ($\gamma$,$X_{i}$) channels can be approximated by the measured yields of the charged particle $X$ within acceptable uncertainty. This allows to experimentally obtain the yields of the ($\gamma$,$X_{i}$) channels which dominantly contribute to the total yields of $X$. Using the simulated yields, these energy ranges for each of the seventeen photon-induced reactions are derived.
Furthermore, the energy spectra of the ($\gamma$,$X_{i}$) channels with $0\leq i\leq 10$ are simulated for each considered reaction, with the incident $\gamma$-beam energies in the respective energy range as derived before. Based on the energy spectra, the identification of the individual dominant ($\gamma$,$X_{i}$) channels is discussed.
It becomes evident that measurements of the photon-induced reactions with charged-particle emissions considered in this work are feasible with the VEGA+ELISSA system and will provide knowledge useful for nuclear astrophysics.
\end{abstract}

\maketitle

\section{Introduction}
\label{sec:intro}

It has been suggested that 32 proton-rich isotopes ($p$ nuclides) between selenium and
mercury originate from stellar explosions
\cite{B2FH,arnould1969,arnould2003p,arnould2007r,kappeler2011s,rauscher2013constraining}. The majority of these nuclides can be
produced by photodisintegration of pre-existing seed nuclei at the high temperatures achieved in supernovae
\cite{woosley1978p,rayet1990p,fujimoto2003p,rapp2006sensitivity,hayakawa2008empirical,travaglio2014radiogenic,
pignatari2016production,rauscher2016uncertainties,nishimura2018impacts,mnrasp}. This is called the $\gamma$ process and involves the
photon-induced release of neutrons, protons, and $\alpha$ particles from nuclei not only at stability but also several mass units
off stability towards the proton-rich side of the nuclear chart. Proton capture may also contribute for the lighter $p$-nuclides. In any case, the required temperatures of the stellar plasma to affect the abundances of $p$-nuclides are in the range of $1.5-3.5$
GK. At such temperatures a significant fraction of all nuclei are found in thermally excited states
\cite{fowler74,rauscher2011gs,rauscher2012sensitivity,rauscherbook}. Therefore, an experimental
constraint of the astrophysical reaction rates for these reactions faces two challenges: dealing with unstable targets and having to
determine reactions on excited nuclear states. Astrophysical simulations use theoretical reaction rates for all of these reaction
rates, with few exceptions for the lighter nuclides.

It has to be realized that there is a reciprocity relation between forward and reverse reactions and that astrophysical
photodisintegration rates are directly related to astrophysical capture rates (see, e.g., \cite{rauscherbook}). Accounting for
reactions on excited nuclear states is a prerequisite that this reciprocity relation holds. It has been shown that significantly fewer excited states contribute in the capture direction whereas they dominate over ground-state transitions by a few orders of magnitude in
photodisintegration rates \cite{rauscherpath}. Therefore a direct measurement of a photodisintegration reaction in the laboratory,
with the target nucleus being in the ground state, cannot constrain the astrophysical rate unless it is possible to assume a correlation between transitions on the ground state and on excited states \cite{rauscherformalism}.

Nevertheless, photodisintegration experiments can help to obtain information on nuclear properties entering the reaction models
used for the prediction of the astrophysical rates. Significant efforts have been undertaken to study the photon-strength function,
which is not only of interest for astrophysics but also for nuclear structure studies \cite{pietralla2019photonuclear}. So far, the
photodisintegration measurements have been conducted using the laser-Compton scattering $\gamma$-ray source
\cite{utsunomiya2003cross,shizuma2005photodisintegration,utsunomiya2008m,utsunomiya2019}, bremsstrahlung radiation
\cite{mohr2000experimental,vogt2001measurement,sonnabend2004systematic,hasper2008investigation,
hasper2009investigation,nair2007photodisintegration,nair2010photodisintegration,erhard2010experimental} and Coulomb dissociation
\cite{ershova2011coulomb,ershova2012coulomb}.

Additionally, it has been suggested that photodisintegration experiments can also probe particle emission to excited states in the final nucleus and thus provide information on the transitions needed to calculate astrophysical capture rates \cite{rauscherA,rauscherB,rauschernews}. This allows to better constrain the reaction models by being able to test calculations, e.g., by using different optical potentials in the computation of the relevant transitions and comparing them to the data \cite{lan2018determination}. Since fewer excited states contribute to the capture rate than the photodisintegration rate, one can place significantly stronger limits on the astrophysically relevant rates using the former rate versus the latter rate. The photodisintegration rate can then be obtained by applying the principle of detailed balance on the improved capture rate.

Despite of the previous efforts, the available experimental information concerning charged-particle emission to final excited states still is too limited to cover all the detailed nuclear transitions needed for the calculations. The Variable Energy Gamma-ray (VEGA) system, which is being constructed at the Extreme Light Infrastructure - Nuclear Physics (ELI-NP) facility, will be able to address this issue and to provide additional data to test and improve theoretical models. Quasi-monochromatic, high-intensity $\gamma$-rays up to 19.5 MeV can be delivered by VEGA, which will unlock new research opportunities on the studies of nuclear physics \cite{NIMB2015ELINP,EPL2017ELINP}, nuclear astrophysics~\cite{ELIcombi2016,ELISSA2016,jinst2017,lan2018determination}, as well as industrial and medical applications~\cite{luo2016production,zhu2016,luo2016estimates,bobeica2016radioisotope}. The silicon strip array at ELI-NP (ELISSA) is specifically constructed for charged-particle detection in photo-nucleation reactions. It is well suited to experimentally study ($\gamma$,p) and ($\gamma$,$\alpha$) reactions at energies corresponding to $\gamma$-process temperatures. It is planned to specifically investigate the cross sections of the ($\gamma$,p$_{i}$) and ($\gamma$,$\alpha_{i}$) reaction channels that proceed via the transitions from the intermediate state of the compound nucleus to different $i$-th final states in the residual nucleus with proton or $\alpha$-particle emissions. Thus, it will not only be possible to better constrain transitions to/from excited states in astrophysical captures in general but specifically also the prediction of capture rates on unstable target nuclides can be treated in this way as long as the compound nucleus (which is the target nucleus in photodisintegration experiments) is stable.

In this paper, the feasibility of studying photon-induced reactions of astrophysical interest with VEGA and ELISSA is
investigated.
We focus on proton emission from seventeen targets ($^{29}$Si, $^{47}$Ti, $^{56}$Fe, $^{73}$Ge, $^{74}$Se, $^{84}$Sr, $^{91}$Zr, $^{95}$Mo, $^{96, 9s8}$Ru, $^{102}$Pd, $^{106}$Cd, $^{115, 117, 119}$Sn, $^{132}$Ba, and $^{142}$Nd) and $\alpha$-particle emission from seventeen targets ($^{50}$V, $^{67}$Zn, $^{87}$Sr, $^{107}$Ag, $^{113, 115}$In, $^{119}$Sn, $^{123, 125}$Te, $^{149, 154}$Sm, $^{155, 156, 157, 158, 160}$Gd, and $^{207}$Pd). For these seventeen ($\gamma$,p) and seventeen ($\gamma$,$\alpha$) reactions, the cross sections for the transitions leading to the ground state and the first ten excited states in the residual nucleus (namely ($\gamma$,p$_{i}$) and ($\gamma$,$\alpha_{i}$) with $0\leq i\leq 10$) are respectively computed in Section \ref{sec:calcs}, in which the sensitivities of the nuclear ingredients to the calculations are illustrated as well.
In Section \ref{sec:feasibility}, based on the features of VEGA and ELISSA at ELI-NP, the results of GEANT4 simulations for the measurements of the reactions are presented, and the feasibility of measuring the individual ($\gamma$,p$_{i}$) and ($\gamma$,$\alpha_{i}$) reaction channels is explicitly investigated. The summary is given in Section \ref{sec:summary}.

\section{Theory and Calculation}
\label{sec:calcs}
\subsection{Reaction model and astrophysical reaction rate}

The compound nucleus contribution (CNC) is the dominant contribution to the reaction cross sections for the production of $p$
nuclides. It is well described by the Hauser-Feshbach reaction model \cite{hauser1952inelastic}. This model relies on the
fundamental Bohr hypothesis that the reaction occurs by means of the intermediary formation of a compound nucleus that can reach a
state of thermodynamic equilibrium. It is valid if the formation and decay of the compound nucleus are independent
\cite{holmes1976,cowan1991r,goriely1996new,rauscher1997nuclear,rauscherbook}. The Hauser-Feshbach model averages over a large
number of resonances in the formation of a compound nucleus and therefore can be applied if the nuclear level density (NLD) in the
compound nucleus is sufficiently high at the compound formation energy, i.e., the excitation energy at which the compound nucleus is
formed. This is always fullfilled for cross sections relevant to the nucleosynthesis of the $p$ nuclides.

In the following, $\sigma_{A^{i}+\gamma\rightarrow B^{j}+b}^{CNC}(E)$ denotes the reaction cross section for the reaction
$A^i+\gamma=B^j+b$ ($b$ = proton or $\alpha$-particle), i.e., photodisintegration of the $i$-th state in nucleus $A$ leading to
particle emission and leaving the final nucleus $B$ in its $j$-th state. Each state is characterized by a spin $S^{i}_{A}$, a
parity $\pi^{i}_{A}$ and an excitation energy $E^{i}_{A}$ for the target $A$ and similarly for the residual nucleus $B$ with the
spin $S^{j}_{B}$, the parity $\pi^{j}_{B}$ and the excitation energy $E^{j}_{B}$.
In the Hauser-Feshbach model, the cross section $\sigma_{A^{i}+\gamma\rightarrow B^{j}+b}^{CNC}(E)$ from $i$-th state in
$A$ to $j$-th state in $B$ is given by (e.g.~\cite{holmes1976,rauscherpath,rauscherbook,goriely2008improved,xu2014systematic})
\begin{eqnarray}
&&\sigma_{A^{i}+\gamma\rightarrow B^{j}+b}^{CNC}(E_{\gamma})=\frac{\pi}{k^2}\sum^{l_{max}+S_{A^{i}}+S_{\gamma}}_{J=mod(S_{A^{i}}+S_{
\gamma},1)}\sum^{1}_{\Pi=-1}\nonumber\\
&&\frac{2J+1}{(2S_{A^{i}}+1)(2S_{\gamma}+1)}
\sum^{J+S_{A^{i}}}_{\lambda=|J-S_{A^{i}}|}
\sum^{\lambda+S_{\gamma}}_{l_{A}=|\lambda-S_{\gamma}|}
\sum^{J+S_{B^{j}}}_{J_{b}=|J-S_{B^{j}}|}
\sum^{J_{b}+S_{b}}_{l_{f}=|J_{b}-S_{b}|}\nonumber\\
&&\delta^{\pi}_{C_\gamma}\delta^{\pi}_{C_b}\frac{\langle T^{J}_{C_\gamma,l_{A},\lambda}(E_\gamma)\rangle
\langle T^{J}_{C_b,l_{f},J_{b}}(E_b)\rangle}
{\sum_{Clj}\delta^{\pi}_{C}\langle T^{J}_{Clj}(E_{C})\rangle}
W^{J}_{C_{\gamma}l_{A}\lambda C_{b}l_{f}J_{b}},
\label{eqHF2}
\end{eqnarray}
\noindent
where $E_{\gamma}$ is the incident energy of $\gamma$-photon; $k$ the wave number of the relative motion; $l_{max}$ the maximum
value of the relative orbital momentum; $J$ and $\Pi$ the total angular momentum and parity of the compound nucleus; $S_{A^{i}}$
the spin of target $A^{i}$, $S_{\gamma}$ the spin of photon, $\lambda$ the multi-polarity of the photon (total angular momentum of
photon), $l_{A}$ the relative orbital momentum of target $A^{i}$ and photon; $S_{B^{j}}$ the spin of residual nucleus $B^{j}$,
$S_{b}$ the spin of emitted particle (proton or $\alpha$ particle here), $J_{b}$ the total angular momentum of emitted particle,
$l_{f}$ the relative orbital momentum of the residual nucleus $B^{j}$ and emitted particle; $E_{b}$ the energy of emitted particle;
$C_{\gamma}$ the channel label of the initial system ($A^{i}$+$\gamma$) designated by $C_{\gamma}$=($\gamma$, $S_{\gamma}$,
$E_{\gamma}$, $E_{A^{i}}$, $S_{A^{i}}$, $\pi_{A^{i}}$); $C_{b}$ the channel label of the final system ($B^{j}$+$b$) designated by
$C_{b}$=($b$, $S_{b}$, $E_{b}$, $E_{B^{j}}$, $S_{B^{j}}$, $\pi_{B^{j}}$); $\delta^{\pi}_{C_{\gamma}}$=1 if
$\pi_{A^{i}}$$\pi_{\gamma}$$(-1)^{l_{A}}$=$\Pi$ and 0 otherwise; $\delta^{\pi}_{C_{b}}$=1 if
$\pi_{B^{j}}$$\pi_{b}$$(-1)^{l_{f}}$=$\Pi$ and 0 otherwise; $\pi_{\gamma}$ the parity of photon, $\pi_{b}$ the parity of emitted
particle; $\langle T \rangle$ the transmission coefficient; $\sum_{Clj}\delta^{\pi}_{C}\langle T^{J}_{Clj}(E_{C})\rangle$ the sum of the
transmission coefficients for all possible decay channels $C$ of the compound nucleus; and $W$ the width fluctuation correction
factor for which different approximate expressions are described and discussed in Ref.~\cite{hilaire2003comparisons}. In particular,
the transmission coefficient for particle emission is determined by the optical model potentials between the two interacting
particles, while the photon transmission coefficient depends on the photon strength function folded with the number of available
final states and thus also on the level density.

When the incident energies increase approximately above the neutron separation energy, the residual nuclei formed after the first
binary reaction are populated with enough excitation energy to enable further decay by particle emission. This refers to multiple
emission, which can be conventionally and sufficiently described by the mechanism of multiple compound Hauser-Feshbach decay in
the energy range from the neutron separation energy up to several tens of MeV. In this mechanism, after the first binary reaction
the excited state $j$ in the residual nucleus emits secondary and further particles if energetically permitted. The population of
this excited state is given by a probability distribution of the Hauser-Feshbach decay that can be completely determined by the
formula of the binary reaction. In practice, the initial compound nucleus energy for the binary emission is replaced by the
excitation energy of the initial nucleus for the multiple compound emission and the calculation of the emission from the excited
state is the same as the calculation of the binary reaction as shown in Eq.\ (\ref{eqHF2}).

Coming back to the simple binary case, the cross section measured in a laboratory experiment, with the target nuclei being in their
ground states ($i=0$), is given by
\begin{eqnarray}
\sigma^{CNC}(E)=\sum_j \sigma_{A^{0}+\gamma\rightarrow B^{j}+b}^{CNC}(E)\quad,
\label{eqHF1}
\end{eqnarray}
summing over all energetically accessible final states in nucleus $B$ (including the ground state with $j=0$). This shows that
particle-transmission coefficients can be probed by photodisintegration experiments.
Reactions in an astrophysical plasma at elevated temperatures, however, have to be described differently. In addition to the ground
state, also excited states are populated and therefore transitions from these excited target states also have to be included,
weighted by the temperature-dependent population $P^A_i(T)$ of the excited states in nucleus $A$. The astrophysical
photodisintegration rate $\lambda^*_\gamma=n_A L^*_\gamma$ ($n_A$ being the number density of target nuclei in the plasma) then
includes a weighted sum of rates on the individual target states \cite{fowler74}, with $L^*_\gamma$ being
\begin{equation}
\label{eq:photorate}
L^*_\gamma(T)=C_\gamma \sum_i P^A_i(T) \int\limits_0^\infty  \sum_j \left\{\sigma_{A^i+\gamma\rightarrow
B^{j}+b}^{CNC}(E_\gamma^i)\right\} {E_\gamma^i}^2 \left[\mathrm{e}^{E_\gamma^i/(k_\mathrm{B}T)}-1\right]^{-1}
\,\mathrm{d}E_\gamma^i\quad,
\end{equation}
with the plasma temperature $T$, the photon energy $E_\gamma^i$, the Boltzmann constant $k_\mathrm{B}$, and a proportionality
constant $C_\gamma$. Here the photon distribution is described by the black-body Planck spectrum at the given temperature $T$. The population of the excited states is given by the Boltzmann factors $P^A_i(T)$. A similar equation can be
found for capture reactions,
\begin{equation}
\label{eq:caprate}
R^*=F(T)\sum_j P^B_j(T) \int\limits_0^\infty  \sum_i \left\{\sigma_{B^{j}+b\rightarrow
A^i+\gamma}^{CNC}(E^j)\right\} E^j
\mathrm{e}^{-E^j/(k_\mathrm{B}T)} \,\mathrm{d}E^j\quad,
\end{equation}
with the population factors $P^B_j(T)$, the c.m.\ energy $E^j$ of the projectiles $b$, the temperature-dependent factor $F(T)$, and
the capture rate being $r^*=n_B n_b R^*$. Note that the reactivity $R^*$ plays the same role for the capture rate as $L^*_\gamma$
in the calculation of the photodisintegration rate but has different dimensions. Further note that the energy $E^i_\gamma$ and
$E^j$ in the integration are given relative to the excitation energy of the $i$-th state in $A$ and $j$-th state in $B$,
respectively. This implies that the integral in each summand has a shifted energy scale relative to the other terms in the sum. This
makes it hard to see that, in fact, the sum in Eq.\ (\ref{eq:photorate}) includes many more states up to a higher excitation energy
than the sum in Eq.\ (\ref{eq:caprate}) when the $Q$ value of the capture reaction is positive. It is more readily seen when
mathematically transforming the sum of integrals in Eq.\ (\ref{eq:caprate}) to a single integral (the complete derivation is given
in \cite{rauscherbook}),
\begin{equation}
\label{eq:singleintegral}
R^*=F'(T) \frac{1}{G_0^B(T)}\int\limits_0^\infty \sum_j \sum_i \frac{2J_j^B+1}{2J_0^B+1}\frac{E-E_j^B}{E}
\sigma_{B^{j}+b\rightarrow
A^i+\gamma}^{CNC}(E-E_j^B) E
\mathrm{e}^{-E/(k_\mathrm{B}T)} \,\mathrm{d}E\quad.
\end{equation}
Here, the cross section is evaluated at an energy $E-E_j^B$ and cross sections at zero or negative energy (i.e., for
$E-E_j^B\leq0$) are set to zero \cite{fowler74}. Excitation energy and spin of excited states in nucleus $B$ are denoted by $E_j^B$
and $J_j^B$, respectively, and $G_0^B(T)$ is the nuclear partition function normalized to the ground state spin. A similar equation
can be found for the photodisintegration rate. The two expressions can be directly compared by realizing that the sum over
excited states $i$ in nucleus $A$ must contain many more terms than the sum over states $j$ in nucleus $B$ because
the excitation energy of the compound state is $E^C=E+Q_\mathrm{Bb}$ when $E$ is the c.m.\ energy of the projectile in a capture
reaction.
Thus, a larger range of thermally excited states is contributing to the photodisintegration rates. This underlines the fact that
measuring $\sigma^{CNC}(E)$ as given in Eq.\ (\ref{eqHF1}), which only includes transitions from the ground state of the target
nucleus, only determines a tiny contribution to the total astrophysical photodisintegration rate. On the other hand, such a
measurement can still obtain information on the particle transitions appearing in the sum over $j$ in Eq.\
(\ref{eq:singleintegral}).

A reciprocity relation between forward and reverse rates can be found, \cite{rauscherbook}
\begin{equation}
L^*_\gamma = R^* \mathcal{F}(T) \mathrm{e}^{-Q_{Bb}/(k_\mathrm{B}T)}\quad,
\end{equation}
where $Q_{Bb}$ is the reaction $Q$-value of the capture reaction and $\mathcal{F}(T)$ is a temperature-dependent factor. This
relation allows to compute the photodisintegration rate when the capture rate is known and vice versa. Note, however, that it only
applies when the contribution of thermally excited states in the respective target nuclides are taken into account. This means that
this relation only applies to proper astrophysical reaction rates as computed from Eqs.\ (\ref{eq:photorate}) and
(\ref{eq:caprate}) but not to rates obtained from simply integrating the laboratory cross section (Eq.\ \ref{eqHF1}), unless
the contribution of excited target states is negligible and the ground-state contribution dominates. Ground-state contributions to
astrophysical reaction rates are given in \cite{rauscher2012sensitivity}. It is found that for nuclides in the vicinity of the $p$
nuclides, the ground-state contributions become negligible at $\gamma$-process temperatures. The ground-state contributions are
much larger for capture reactions, as explained above, but still the contribution of the excited states cannot be neglected in many
cases. Therefore studying charged-particle emission by photodisintegration in the laboratory offers a way to better constrain the
calculated excited state contribution in astrophysical capture rates. In this work, we aim to apply such an approach for
improving the rates implemented in astrophysical simulations of the $\gamma$ process by obtaining data on photon-induced particle
emission from compound nuclear states.

The typical temperature of the astrophysical plasma is $1.5-3.5$ GK for $\gamma$-process, corresponding to the thermal energies of $kT$ = $170-260$ keV ($k$ = 86.173 keV/GK is the Blotzmann constant). The peak of the Planck photon distribution is at 1.5 $kT$ \cite{photon1,photon2,photon3}, which leads to the most probable photon energy to be in the range of $260-390$ keV. Such energy range is mostly covered by the excitation energies of the low-lying states (\textit{e.g.} below 500 keV) in the residual nuclei that are produced from the photodisintegration studied in this paper. The relevant contributions to the integral in Eq.\ (\ref{eq:singleintegral}) at $\gamma$-process temperatures stem from an energy window of about $2-4$ MeV for protons and $5-10$ MeV for $\alpha$-particles \cite{rauscher2012sensitivity,rauscher2010windows}. This means that these are the most important c.m.\ energies for protons and $\alpha$-particles in the astrophysical plasma. Adding the reaction $Q$-value to these ranges yields the compound excitation energy leading to emission of charged particles in this important energy range. Studying the particle-transmission coefficients as obtained from Eq.\ (\ref{eqHF2}) from compound states at these excitation energies provides the best information for the astrophysical application. This is not always possible experimentally due to low charged-particle yields. A measurement at a range of higher energies, nevertheless, may also be used to test the prediction of transmission coefficients and their energy dependence.

\begin{figure*}
\hspace*{-0.2cm}
\centering
\vspace*{-0cm}
\includegraphics[width=17cm,clip]{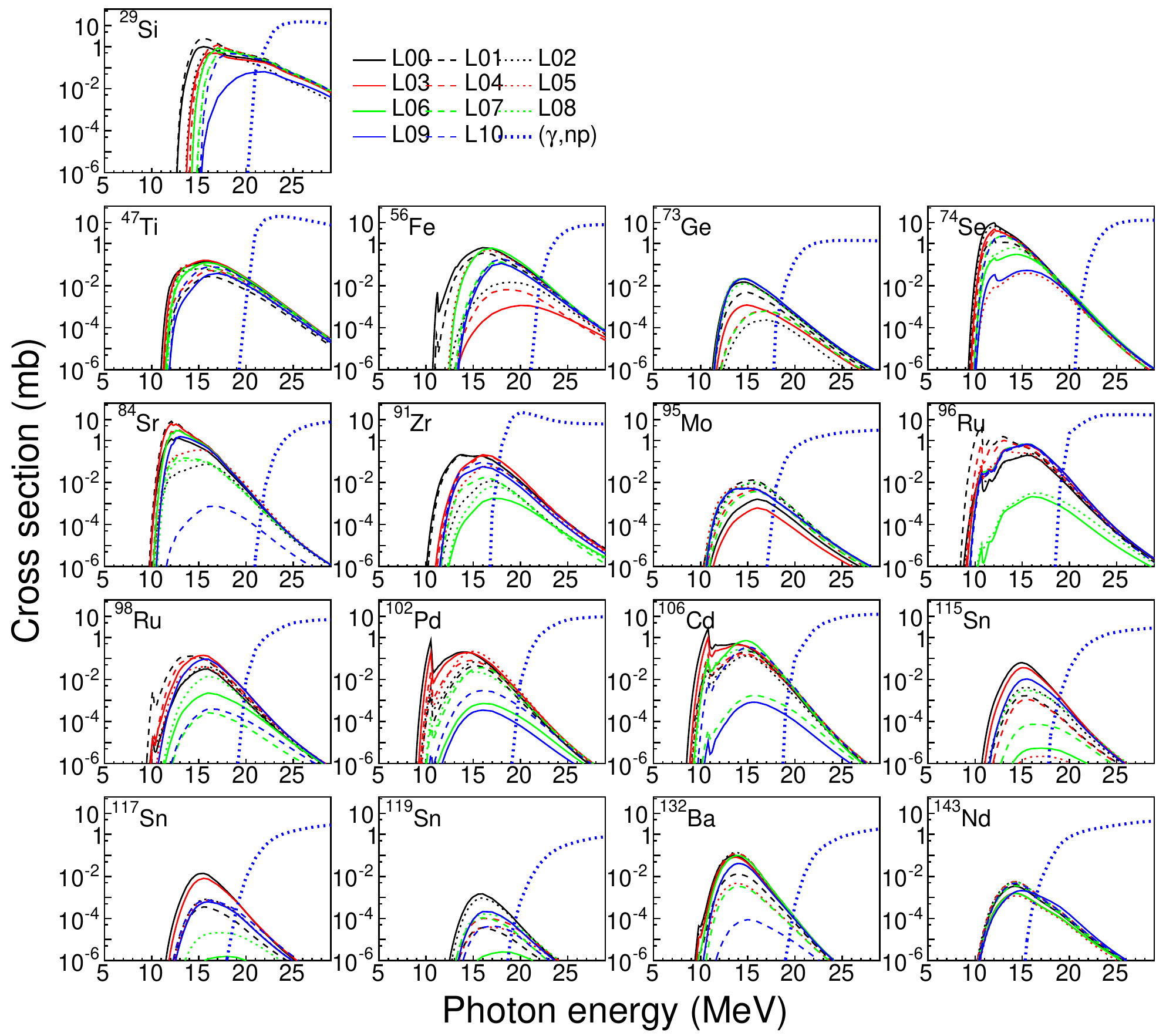}
\caption{\label{cs-gp} The cross sections of the ($\gamma$,p$_{i}$) channels with $0\leq i\leq 10$ and the ($\gamma$,$np$) channel
on seventeen targets of $^{29}$Si, $^{47}$Ti, $^{56}$Fe, $^{73}$Ge, $^{74}$Se, $^{84}$Sr, $^{91}$Zr, $^{95}$Mo, $^{96, 98}$Ru,
$^{102}$Pd, $^{106}$Cd, $^{115, 117, 119}$Sn, $^{132}$Ba and $^{142}$Nd. The ($\gamma$,p$_{i}$) indicates the reaction with proton
emission proceeding on the $i$-th final state ($i$=0 for the ground state) in the residual nucleus. All the results are calculated
by TALYS 1.9.}
\end{figure*}

\begin{figure*}
\hspace*{-0.2cm}
\centering
\vspace*{-0cm}
\includegraphics[width=17cm,clip]{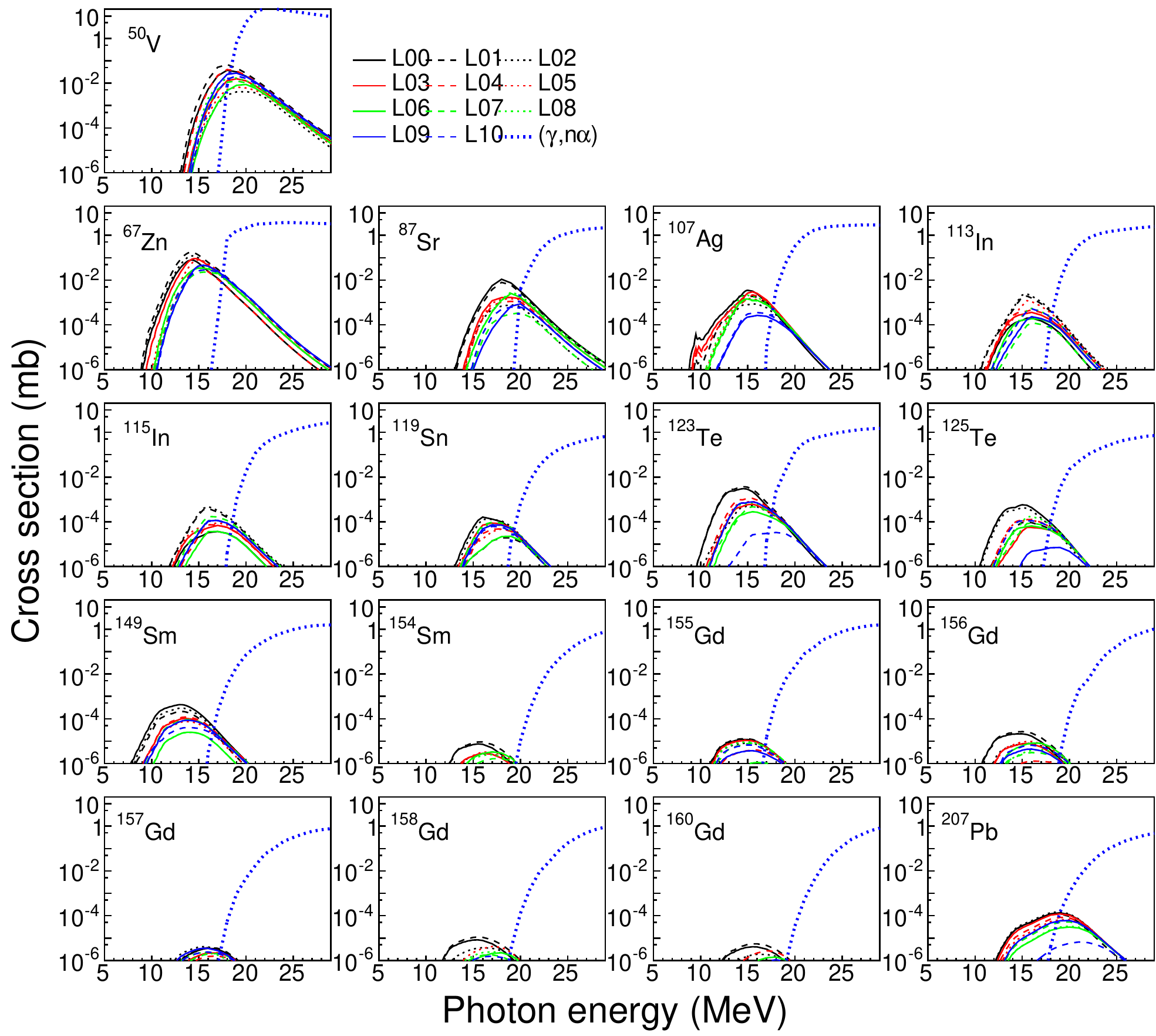}
\caption{\label{cs-ga}
The cross sections of the ($\gamma$,$\alpha_{i}$) channels with $0\leq i\leq 10$ and the ($\gamma$,$n$$\alpha$) channel on seventeen
targets of $^{50}$V, $^{67}$Zn, $^{87}$Sr, $^{107}$Ag, $^{113, 115}$In, $^{119}$Sn, $^{123, 125}$Te, $^{149, 154}$Sm, $^{155, 156,
157, 158, 160}$Gd and $^{207}$Pd. The ($\gamma$,$\alpha_{i}$) indicates the reaction with $\alpha$-particle emission proceeding on
the $i$-th final state ($i$=0 for the ground state) in the residual nucleus. All the results are calculated by TALYS 1.9.}
\end{figure*}

\subsection{Photodisintegration cross sections}
\label{sec-cs}

Among capture reactions shown to have considerable contributions of excited states to the astrophysical rate (as identified in
\cite{rauscher2012sensitivity}) we select cases best suited for measuring photon-induced emission to their excited states at the
VEGA+ELISSA system. The photon-induced reactions with proton emission on 17 target nuclides are considered in the present
study: $^{29}$Si, $^{47}$Ti, $^{56}$Fe, $^{73}$Ge, $^{74}$Se, $^{84}$Sr, $^{91}$Zr, $^{95}$Mo, $^{96, 98}$Ru, $^{102}$Pd,
$^{106}$Cd, $^{115, 117, 119}$Sn, $^{132}$Ba, and $^{142}$Nd. Furthermore, photon-induced $\alpha$-particle emission on 17
targets are considered: $^{50}$V, $^{67}$Zn, $^{87}$Sr, $^{107}$Ag, $^{113, 115}$In, $^{119}$Sn, $^{123, 125}$Te, $^{149, 154}$Sm,
$^{155, 156, 157, 158, 160}$Gd, and $^{207}$Pd.

For the selected photon-induced reactions, the cross sections of ($\gamma$,p$_{i}$) and ($\gamma$,$\alpha_{i}$) channels, as well as the multiple emission channels ($\gamma$,$np$) and ($\gamma$,$n$$\alpha$), are computed with TALYS 1.9. The nuclear structure ingredients used in the TALYS calculations are as follows. The nuclear masses are taken from the Atomic Mass Evaluation 2016 (AME2016) \cite{ame2016} whenever available, while the HFB-27 nuclear masses~\cite{goriely2013further} are taken into account when the AME2016 mass data are not available. The discrete experimental levels compiled in RIPL-3 library \cite{capote2009ripl} and the continuum level spectrum represented by the nuclear level densities (NLDs) are both considered in the calculations. The NLDs are taken from predictions in the microscopic HFB model plus a combinatorial approach \cite{goriely2008nld} that can well reproduce the low-lying cumulative number of the experimental levels. The photon strength functions obtained from the HFB plus quasiparticle random phase approximation (QRPA) \cite{goriely2004psf} are used to calculate the electromagnetic transmission coefficients for the photon channel.
The optical model potentials (OMPs) are employed to determine the transmission coefficient for the particle (proton and
$\alpha$-particle) channels. Specifically, the JLM microscopic optical potential \cite{Goriely2007pot} and the M3Y double folding dispersive potential \cite{demetriou2002} are used for the proton and the $\alpha$-particle
channels, respectively.

The calculated results for the proton emission are shown in Figure~\ref{cs-gp}. The highest ($\gamma$,p$_{i}$) cross section for $^{56}$Fe, $^{96}$Ru and $^{102}$Pd reaches about 1 mb, while that for $^{119}$Sn is only a few of 10$^{-3}$ mb. At $E_{\gamma}< 16$ MeV, the ($\gamma$,p$_{i}$) cross sections for the transitions to the ground state and the first two excited states ($i=0$, 1 and 2) are dominant for most target nuclides. However, when the incident $E_{\gamma}$ is above 16 MeV, the ($\gamma$,p$_{i}$) cross sections with $i>2$ become relevant. This means that the transitions to the final states with higher excited energies in the residual nucleus are significant when the incident $E_{\gamma}$ increases. Note that for the targets of $^{96}$Ru and $^{98}$Ru, the cross section of ($\gamma$,p$_{1}$) for the transition to the first excited state, rather than the ground state, is dominant at $E_{\gamma}< 16$ MeV. Furthermore, compared to the cross section of single proton emission, the cross section of multiple emission via ($\gamma$,np) is negligible for the target nuclei $^{29}$Si, $^{56}$Fe, $^{91}$Zr and $^{96}$Ru at $E_{\gamma}< 20$ MeV, but remarkably even dominant for $^{98}$Ru, $^{102}$Pd, $^{106}$Cd, $^{115}$Sn, $^{117}$Sn and $^{119}$Sn when $E_{\gamma}$ increases above 18 MeV.

The results for the $\alpha$-particle emission are shown in Figure~\ref{cs-ga}. At $E_{\gamma}< 18$ MeV, the ($\gamma$,$\alpha_{i}$) cross sections for the transitions to the ground and the first two excited states ($i=0$, 1 and 2) in the residual nucleus are dominant. For the $^{50}$V target, the ($\gamma$,$\alpha_{1}$) cross section is prominent at $E_{\gamma}< 20$ MeV. The cross sections of ($\gamma$,n$\alpha$) are comparable to those of ($\gamma$,$\alpha_{i}$) for $^{50}$V, $^{87}$Sr, $^{123}$Te, and $^{149}$Sm when the incident energy $E_{\gamma}$ reaches 19 MeV.


The sensitivity of the calculated cross sections to the nuclear structure ingredients have been systematically studied in our previous work \cite{lan2018determination}. It is found that the total cross sections of ($\gamma$,p) and ($\gamma$,$\alpha$) are dramatically influenced by the nuclear potentials of the charged particles. In the present study, we briefly investigate how the three kinds of nuclear structure ingredients, $i.e.$, the nuclear level density (NLD), the optical model potential (OMP), and the photon-strength function (SF), impact the contribution of the ($\gamma$,p$_{i}$) or ($\gamma$,$\alpha_{i}$) cross section for a specific $i$-th final state in the residual nucleus to the total cross section of proton or $\alpha$-particle emission. In particular, the cross section ratios of ($\gamma$,p$_{1}$) to ($\gamma$,p$_\mathrm{tot.}$) for $^{96}$Ru and ($\gamma$,$\alpha_{0}$) to ($\gamma$,$\alpha_\mathrm{tot.}$) for $^{123}$Te are calculated using different sets of NLDs, OMPs and SFs available in TALYS.

\begin{figure*}
\hspace*{-0.2cm}
\centering
\vspace*{-0cm}
\includegraphics[width=17cm,clip]{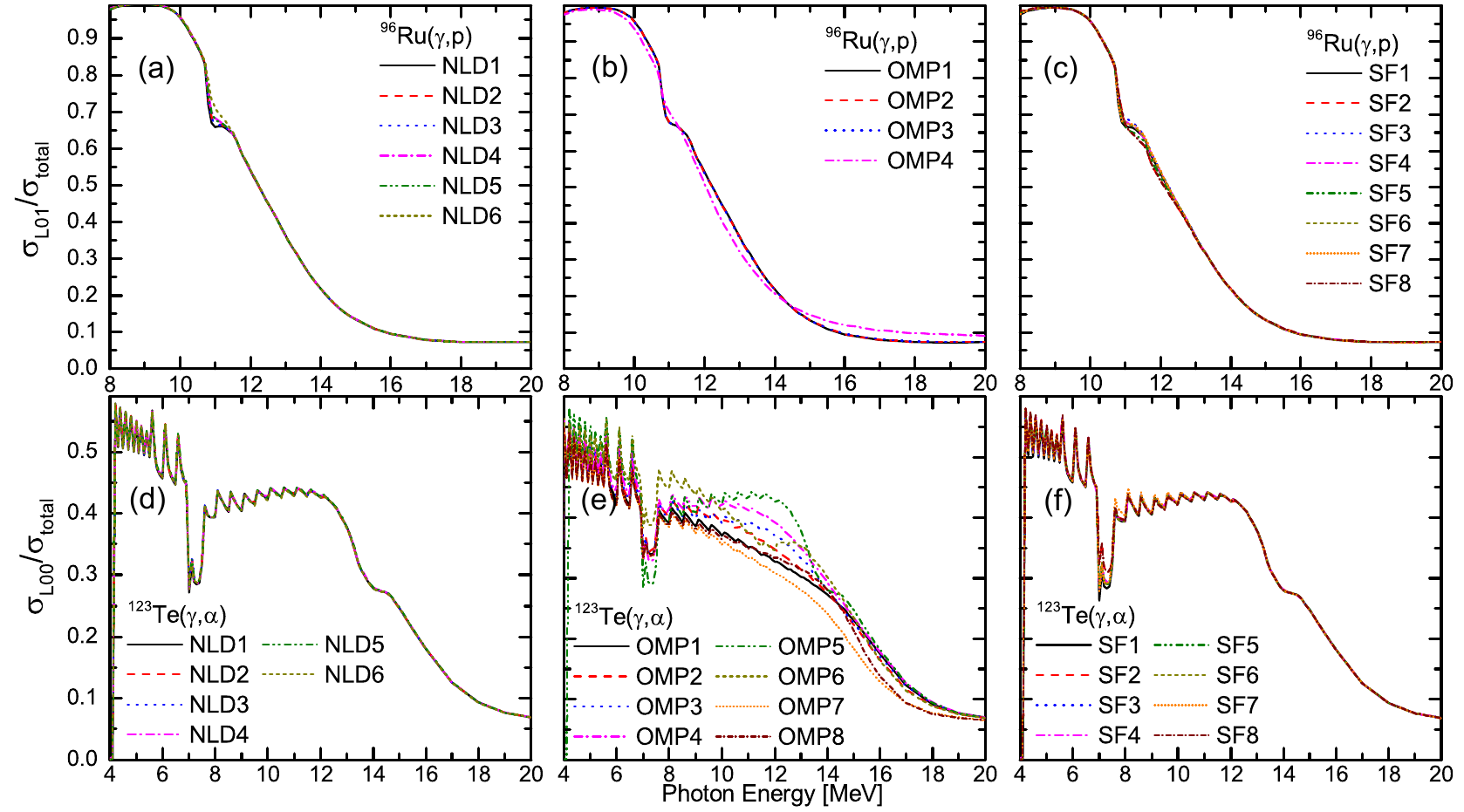}
\caption{\label{sensi} The ratios of the ($\gamma$,p$_{1}$) cross section to the ($\gamma$,p$_\mathrm{tot.}$) cross section for
$^{96}$Ru calculated with six sets of NLDs (a), four sets of OMPs (b) and eight sets of SFs (c) available in TALYS 1.9, and the
ratios of the ($\gamma$,$\alpha_{0}$) cross section to the ($\gamma$,$\alpha_\mathrm{tot.}$) cross section for $^{123}$Te
calculated with six sets of NLDs (d), eight sets of OMPs (e) and eight sets of SFs (f) available in TALYS 1.9. The nuclear
structure ingredients used for the calculations are indexed by numbers and described in the main text.}
\end{figure*}

The results for $^{96}$Ru predicted using six sets of NLDs (1 - Constant temperature Fermi gas model \cite{nld1}, 2 - Back-shifted Fermi gas model \cite{nld2}, 3 - Generalised superfluid model \cite{nld3}, 4 - HFB-Skyrme model \cite{nld4}, 5 - HFB-Skyrme model with combinatorial method \cite{goriely2008nld}, and 6 - Temperature(T)-dependent HFB-Gogny model \cite{nld6}), four sets of OMPs (1 - Wood-Saxon potential \cite{koning2003}, 2 - JLMB potential with HFB-Skyrme matter density \cite{jeukenne1976,jeukenne1977,bauge1998,goriely2013further}, 3 - JLMB potential with HFB-Gogny matter density \cite{jeukenne1976,jeukenne1977,bauge1998,Goriely2009D1M}, and 4 - JLMB potential with HFB-Skyrme matter density plus modified imaginary part \cite{Goriely2007pot}), and eight sets of PSFs (1 - Generalized Lorentzian \cite{Kopecky1990}, 2 - Brink-Axel Lorentzian \cite{Brink1957,Axel1965}, 3 - HFBCS-QRPA model \cite{Goriely2002psf}, 4 - HFB-Skyrme-QRPA model \cite{Goriely2004psf}, 5 - Hybrid model \cite{Goriely1998psf}, 6 - T-dependent HFB-Skyrme-QRPA model \cite{Goriely2004psf,Hilaire2012}, 7 - T-dependent RMF model \cite{Goriely2012psf}, and 8 - HFB-D1M-QRPA model \cite{Goriely2018psf}) are shown in Figure \ref{sensi}(a) - \ref{sensi}(c). The results for $^{123}$Te calculated using six sets of NLDs (same as those of $^{96}$Ru), eight sets of OMPs (1 - global WS potential derived from simple folding approach \cite{Watanabe1958}, 2 - global WS potential for nucleus between O and U \cite{mcfadden1966}, 3 - double folding M3Y real plus WS volume imaginary potential \cite{demetriou2002}, 4 - double folding M3Y real plus WS volume and surface imaginary potential \cite{demetriou2002}, 5 - M3Y-based dispersive model potential \cite{demetriou2002}, 6 - Global WS potential with surface imaginary for $40\leq A\leq 200$ \cite{avrigeanu1994}, 7 - Global WS potential derived from higher energy $\alpha$ scattering \cite{nolte1987}, and 8 - Global WS potential with the extension for lower energies \cite{avrigeanu2014}), and eight sets of PSFs (same as those of $^{96}$Ru) are shown in Figure \ref{sensi}(d) - \ref{sensi}(f).
Note that for such calculations, when a specific nuclear ingredient is changed, other nuclear ingredients generating Figure \ref{cs-gp} and Figure \ref{cs-ga} are kept constant.
It is shown that the calculated ratios are not influenced by the changing of NLDs and SFs, but vary within a maximum
of 10$\%$ caused by different OMPs. Note that the dominance of the two studied contributions shown in Figure \ref{sensi} is not impacted, regardless of the fact that different nuclear ingredients have been used in the calculation.

\section{GEANT4 Simulation}
\label{sec:feasibility}

\subsection{Infrastructure for measurements: VEGA and ELISSA at ELI-NP}

The Extreme Light Infrastructure - Nuclear Physics (ELI-NP) \cite{NIMB2015ELINP} is aiming to use extreme electromagnetic fields for nuclear physics research, which comprises a high power laser system (HPLS) and a variable energy $\gamma$-ray system (VEGA). For the VEGA at ELI-NP, the high-brilliance narrow-bandwidth $\gamma$-beam, produced via the Compton backscattering of a laser beam off a relativistic electron beam, will be delivered, with the spectral density of $5\times10^{3}$ photons$/$s$/$eV, the energies up to 19.5 MeV, and the bandwidth of 0.5$\%$. Thanks to such features of the $\gamma$-beam, ELI-NP could provide unique opportunities to experimentally study the photo-induced reactions of nuclear astrophysics interest.

For the detection of charged particles, at ELI-NP the silicon strip array (ELISSA) has been implemented and tested \cite{ELISSA2016}. The GEANT4 simulation proves that a barrel configuration of ELISSA is particularly suited as it not only guarantees a very good resolution and granularity but also ensures a compact detection system and a limited number of electronics channels \cite{lattuada2017}. The final design of ELISSA consists of three rings of twelve X3 position sensitive detectors produced by Micron Semiconductor Ltd. \cite{Micron} in a barrel-like configuration, with the assembly of four QQQ3 segmented detectors produced by Micron Semiconductor Ltd. as the end caps of both sides. Such configuration ensures a total angular coverage of 20$\leq$$\theta$$\leq$160 in the laboratory system. The prototype of ELISSA has been constructed and tested at INFN-LNS, and the preliminary experimental results show that the energy resolution is better than 1$\%$ and that the spatial resolution is up to 1 mm \cite{jinst2017,chesnevskaya2018,guardo2017}.

\subsection{Algorithm of simulation}

The measurements of the photon-induced reactions on the seventeen targets with proton emission and the seventeen targets with $\alpha$-particle emission are simulated using the data-based Monte Carlo simulation program
GEANT4-GENBOD~\cite{luo20114d,luo2017implementation}. The ($\gamma$,p$_{i}$) and ($\gamma$,$\alpha_{i}$) channels (0 $\leq$ $i$ $\leq$ 10), as well as the multiple emission channels ($\gamma$,np) and ($\gamma$,n$\alpha$), are taken into account in the simulation, and correspondingly the calculated cross sections in Sec. \ref{sec-cs} are incorporated as the inputs for the simulation. The features of the $\gamma$-beam facility at ELI-NP with the photon intensity of $10^{4}$ photons$/$s$/$eV and the energy bandwidth of 1$\%$ are taken into account in the simulation, and the configuration of ELISSA is implemented. A double-layer target, consisting of a 10-$\mu$m thick target facing the $\gamma$-beam and a 0.266-$\mu$m thin carbon backing, is used in the simulation.

In order to separate the interesting charged particles produced from the photo-induced reactions, discrimination should be made on the energy spectra of the outgoing particles including electron, proton and $\alpha$-particle. Our previous study \cite{lan2018determination} has suggested that the proton and the $\alpha$-particle produced from the photo-induced reactions can be simultaneously measured, when the incident energy of the $\gamma$-ray ($E_{\gamma}$) is larger than the threshold by 1.0 - 2.0 MeV. Furthermore, it is expected that the pulse shape analysis \cite{duenas2012identification} can be employed to realize the particle identification of proton and $\alpha$-particle. Besides the photonuclear reactions, the incident $\gamma$-beam can also induce Compton effect and pair production in the target, which may influence the detection of the energies of the emitted charged particles. In the present study, Compton effect and pair production are taken into account by invoking the electromagnetic physical process in GEANT4 simulation, and it is indicated that the rate of such background events can be removed by introducing a negligible threshold on the detector \cite{lan2018determination,Matei2020Li7}.

\subsection{Yield simulation and the reactions considered for further studies}
\label{sec-yield}

For the photon-induced reactions on the seventeen targets with proton emission, the simulated yields $Y_{i}^\mathrm{p}$ for ($\gamma$,p$_{i}$) and $Y^\mathrm{np}$ for ($\gamma$,np) are obtained. Likewise, $Y_{i}^{\alpha}$ for ($\gamma$,$\alpha_{i}$) and $Y^{\mathrm{n}\alpha}$ for ($\gamma$,n$\alpha$) are simulated for the seventeen photon-induced reactions with $\alpha$-particle emission. The total yields $Y_\mathrm{total}^\mathrm{p}$ for the proton emission and $Y_\mathrm{total}^{\alpha}$ for the $\alpha$-particle emission are respectively determined by
\begin{eqnarray}
Y_\mathrm{total}^\mathrm{p}= \sum_{i=0}^{10} Y_{i}^\mathrm{p} + Y^\mathrm{np}
\label{eq-yieldp}
\end{eqnarray}
and
\begin{eqnarray}
Y_\mathrm{total}^{\alpha}= \sum_{i=0}^{10} Y_{i}^{\alpha} + Y_{\mathrm{n}\alpha}.
\label{eq-yielda}
\end{eqnarray}

The simulation results reveal that, for the photo-induced reactions with proton emission on twelve targets of $^{29}$Si, $^{56}$Fe, $^{74}$Se, $^{84}$Sr, $^{91}$Zr, $^{96,98}$Ru, $^{102}$Pd, $^{106}$Cd, and $^{115, 117, 119}$Sn, and $\alpha$-particle emission on five targets of $^{50}$V, $^{87}$Sr, $^{123,125}$Te, and $^{149}$Sm, the reaction yields for the transitions to the exited states in the residual nucleus are relevant and even dominant. These seventeen photo-induced reactions with the proton emission on twelve targets and the $\alpha$-particle emission on five targets are further considered in the following simulation studies, because the emphasis of the present paper is to investigate the feasibility of the measurements for the photo-induced reactions proceeding on the excited states in the residual nucleus. Meanwhile, for the remaining seventeen reactions, the simulation results indicate that the reaction yield for the transition to the ground state is prominent all along the incident energies of the $\gamma$-beam, so these seventeen photo-induced reactions are no longer taken into account in the present study.

\begin{figure*}
\hspace*{-0.2cm}
\centering
\vspace*{-0cm}
\includegraphics[width=17cm,clip]{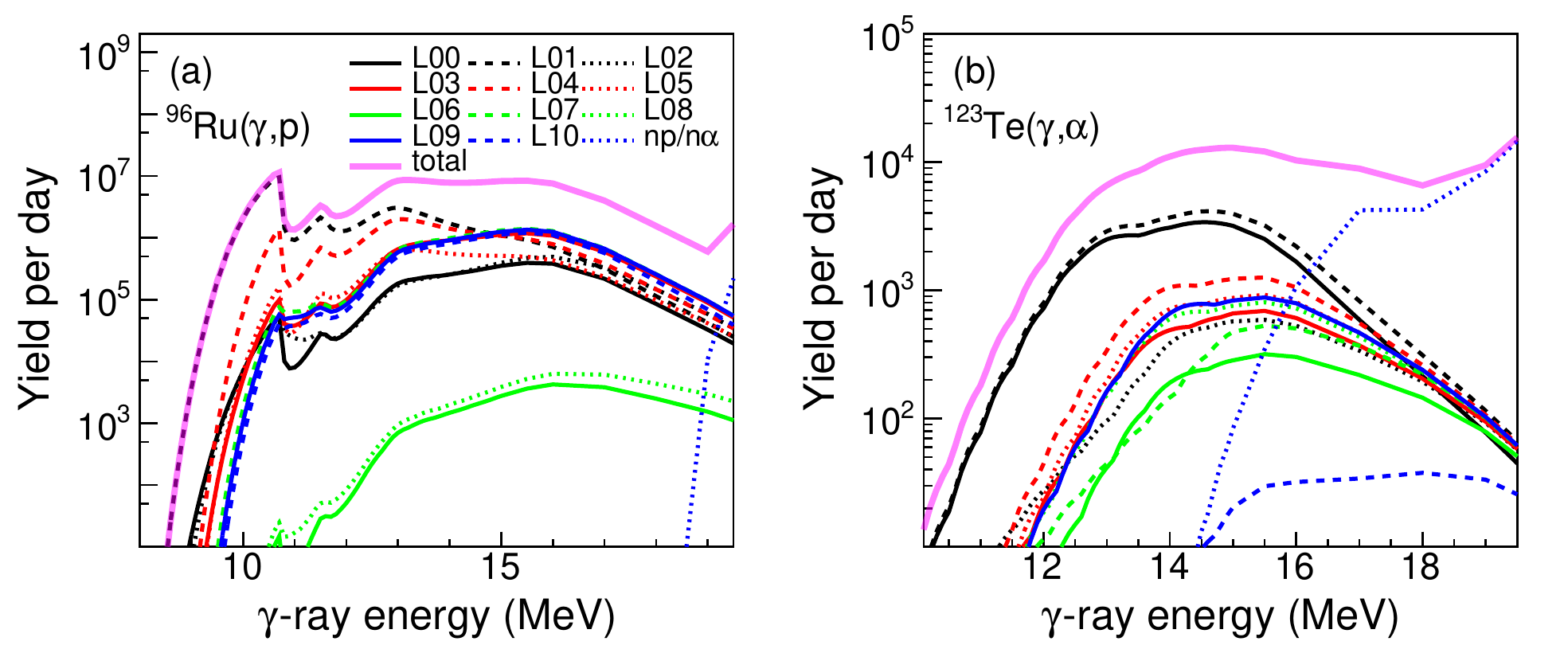}
\caption{\label{totalyield} (a) The simulated yields $Y_{i}^\mathrm{p}$ for $^{96}$Ru($\gamma$,p$_{i}$) with $0\leq i\leq 10$, $Y^\mathrm{np}$ for $^{96}$Ru($\gamma$,np), and $Y_\mathrm{total}^\mathrm{p}$ for
$^{96}$Ru($\gamma$,p$_\mathrm{tot.}$). (b) The simulated yields $Y_{i}^{\alpha}$ for $^{123}$Te($\gamma$,$\alpha_{i}$) with $0\leq i\leq 10$, $Y^{\mathrm{n}\alpha}$ for $^{123}$Te($\gamma$,n$\alpha$), and $Y_\mathrm{total}^{\alpha}$ for $^{123}$Te($\gamma$,$\alpha_\mathrm{tot.}$).}
\end{figure*}

As an example, figure~\ref{totalyield} demonstrates the simulated yields $Y_{i}^\mathrm{p}$, $Y^\mathrm{np}$ and $Y_\mathrm{total}^\mathrm{p}$ for $\gamma$+$^{96}$Ru reaction, and $Y_{i}^{\alpha}$, $Y^{\mathrm{n}\alpha}$ and $Y_\mathrm{total}^{\alpha}$ for $\gamma$+$^{123}$Te reaction. For $^{96}$Ru, it is found that the ($\gamma$,p$_{1}$) yield is dominant up to $E_{\gamma} = 14$ MeV, while the ($\gamma$,p$_{4}$) channel becomes significant with the increase of the incident $\gamma$-beam energy. For $^{123}$Te, the yields of the reactions proceeding on the ground state ($\gamma$,$\alpha_{0}$) and the first excited state ($\gamma$,$\alpha_{1}$) in the residual nucleus are comparable, which are both considerable at the entire energy range of the incident $\gamma$-beam.

\subsection{Feasibility study for the measurement of the ($\gamma$,p$_{i}$) and ($\gamma$,$\alpha_{i}$) channels}
\label{sec-spec}

In order to evaluate the feasibility of the measurements for the considered seventeen photo-induced reactions, it is necessary to estimate the required energies of the incident $\gamma$-beam that can satisfy a minimum measurable limit of the experimental yield.
Our previous study \cite{lan2018determination} has shown that for the measurement of the charged-particles yield, a minimum measurable limit of 100 counts per day is appropriate, considering the details of the ELISSA detector worked out in Ref.~\cite{jinst2017,chesnevskaya2018}.
For each of the seventeen photo-induced reactions considered in the present simulation studies, the total yield of proton or $\alpha$-particle stemming from all the open ($\gamma$,p$_{i}$) or ($\gamma$,$\alpha_{i}$) channels can be measured. Therefore, the minimum measurable limit of 100 counts per day for $Y_\mathrm{total}^\mathrm{p}$ or $Y_\mathrm{total}^{\alpha}$ is taken into account. To meet such criteria, the required energies of the incident $\gamma$-beam, namely $E_\mathrm{req.}$, are deduced for each of the seventeen photo-induced reactions, and the results are listed in Table \ref{t1} (column 4).

\begin{figure*}
\hspace*{-0.2cm}
\centering
\vspace*{-0cm}
\includegraphics[width=17cm,clip]{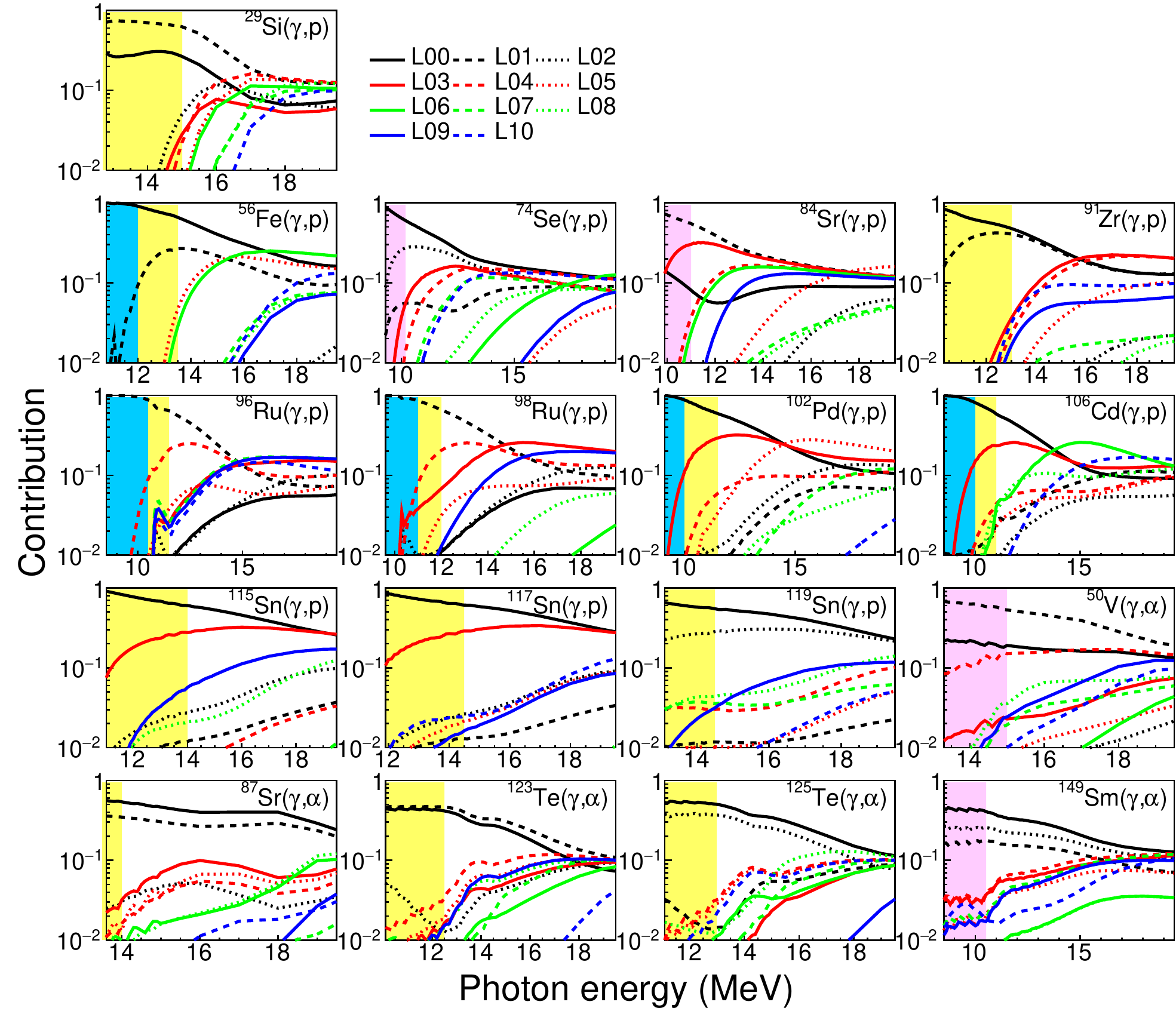}
\caption{\label{contribution} The relative contributions of the ($\gamma$,$X_{i}$) channels ($X$=p or $\alpha$) for each $i$-th
final state ($0\leq i\leq 10$) in the residual nucleus to the total yields of the charged particle $X$,
$Y_{i}^{X}$/$Y_\mathrm{total}^{X}$, for the considered seventeen photo-induced reactions. The blue, yellow and purple bands illustrate
the energy ranges of the incident $\gamma$-beam for performing the measurements, in which one, the sum of two, and the sum of three
($\gamma$,$X_{i}$) channels respectively contribute more than 90$\%$ to the total yield of $X$. See the text for more details.}
\end{figure*}

Furthermore, for the seventeen photo-induced reactions considered in the simulation studies, the relative contributions of the reaction yields for the transitions to different final states in the residual nucleus are investigated. In particular, the yield ratios $Y_{i}^\mathrm{p}$/$Y_\mathrm{total}^\mathrm{p}$ and $Y_{i}^{\alpha}$/$Y_\mathrm{total}^{\alpha}$ for each $i$-th state are calculated from the simulated yields, and the results are shown in Figure~\ref{contribution}. Note that the minimum required energies of the $\gamma$-beam, $E_\mathrm{req.}$, obtained in Sec. \ref{sec-yield} (column 4 in Table \ref{t1}), are set as the lower limit of the X-axis for each reaction.

In Figure \ref{contribution} it can be seen that for all the reactions, when the energy of the incident $\gamma$-beam ($E_{\gamma}$)
is lower, the contributions to the total yields from the transitions to several (usually no more than three) final states in the
residual nucleus are dominant, while no prominent channels contributing to the total yields can be found when $E_{\gamma}$ is
getting higher. Therefore, for each reaction in Figure \ref{contribution}, if the expected experiments are performed within the
specified (narrow) energy ranges of the incident $\gamma$-beam, i.e. [$E_\mathrm{low}$,$E_\mathrm{upper}$], the total reaction
yields
$Y_\mathrm{total}^\mathrm{p}$ or $Y_\mathrm{total}^{\alpha}$ will be dominantly contributed from the sum of the yields of $n$
individual ($\gamma$,p$_{i}$) or ($\gamma$,$\alpha_{i}$) channels, namely $\Sigma_{i\in n}Y_{i}^\mathrm{p}$ or $\Sigma_{i\in
n}Y_{i}^{\alpha}$, with an acceptable relative uncertainty $\Delta$. This can be represented by:
\begin{eqnarray}
(Y_\mathrm{total}^\mathrm{p} - \sum_{i\in n}Y_{i}^\mathrm{p})/Y_\mathrm{total}^\mathrm{p} \le \Delta
\label{eq-delta1}
\end{eqnarray}
or
\begin{eqnarray}
(Y_\mathrm{total}^{\alpha} - \sum_{i\in n} Y_{i}^{\alpha})/Y_\mathrm{total}^{\alpha} \le \Delta.
\label{eq-delta2}
\end{eqnarray}

In practice, the total yields $Y_\mathrm{total}^\mathrm{p}$ and $Y_\mathrm{total}^{\alpha}$, respectively, are straightforwardly
the experimental yields of proton or $\alpha$-particle that can be obtained from the realistic photon-induced measurements.
Therefore, $\Sigma_{i\in n}Y_{i}^\mathrm{p}$ or $\Sigma_{i\in n}Y_{i}^{\alpha}$ can be derived from Eqs. \ref{eq-delta1} and
\ref{eq-delta2} with the given relative uncertainty $\Delta$. This allows us to experimentally determine the yields of the $n$
($\gamma$,p$_{i}$) or ($\gamma$,$\alpha_{i}$) channels that prominently contribute to the total yields with an acceptable
uncertainty. Note that such a procedure is valid only if the energies of the incident $\gamma$-beam for the photo-induced
measurements are in the range of [$E_\mathrm{low}$,$E_\mathrm{upper}$]. Thus it is significant to obtain the energy range
[$E_\mathrm{low}$,$E_\mathrm{upper}$] for each of the seventeen photon-induced reactions.

As far as the present simulation is concerned, the energy range [$E_\mathrm{low}$,$E_\mathrm{upper}$] can be estimated by
satisfying Eqs. \ref{eq-delta1} and \ref{eq-delta2} with the simulated yields ($Y_{i}^{X}$ and $Y_\mathrm{total}^{X}$, $X$=p or
$\alpha$) and the given $n$ and $\Delta$. In the present study, [$E_\mathrm{low}$,$E_\mathrm{upper}$] corresponding to $n$ = 1, 2
and 3 are respectively estimated with $\Delta = 10\%$. This means that, for $n = 1$, a single ($\gamma$,$X_{i}$) channel
contributes more than 90$\%$ to the total yield $Y_\mathrm{total}^{X}$, while for $n =2$ or 3, more than 90$\%$ of the total yield
is contributed from the sum of the yields of 2 or 3 ($\gamma$,$X_{i}$) channels, respectively. For each of the seventeen
photon-induced reactions, the obtained [$E_\mathrm{low}$,$E_\mathrm{upper}$] are listed in table \ref{t1}, and concomitantly, the
identified ($\gamma$,$X_{i}$) channels that dominantly contribute to the total yields in [$E_\mathrm{low}$,$E_\mathrm{upper}$] are
given as well. Meanwhile, the energy ranges [$E_\mathrm{low}$,$E_\mathrm{upper}$] for $n$ = 1, 2 and 3 are respectively illustrated
with the areas in blue, yellow and purple in Figure \ref{contribution} for each reaction.

It can be seen in Figure \ref{contribution} that, for the photo-induced reactions with proton emissions on the targets of $^{56}$Fe,
$^{96}$Ru, $^{98}$Ru, $^{102}$Pd and $^{106}$Cd, the
single ($\gamma$,p$_{0}$) or ($\gamma$,p$_{1}$) channel prominently contributes to the total proton yield. Thus the yields of these
five specific channels, $^{56}$Fe($\gamma$,p$_{0}$), $^{96}$Ru($\gamma$,p$_{1}$), $^{98}$Ru($\gamma$,p$_{1}$),
$^{102}$Pd($\gamma$,p$_{0}$) and $^{106}$Cd($\gamma$,p$_{0}$), can be exclusively determined from the measurements if the energies
of the incident $\gamma$-ray are in the ranges indicated by the blue areas in Figure \ref{contribution}. Note that if the
photonuclear measurements can be performed with the incident $\gamma$-ray in the energy range illustrated by the yellow areas in
Figure \ref{contribution}, more than 90$\%$ of the total yields stem from two specific ($\gamma$,$X_{i}$) ($X$=p or $\alpha$)
channels. Furthermore, for $^{50}$V, $^{74}$Se, $^{84}$Sr and $^{149}$Sm, the sum of the yields of three ($\gamma$,$X_{i}$) channels
contribute at least 90$\%$ of the total yields, when the incident $\gamma$-ray energies are in the range shown by the purple areas in
Figure \ref{contribution}. In this way, the sum of the yields of such two or three specific channels can be directly determined from
the measured yields. However, in order to further derive the yields of each individual channel proceeding on different final
excited states in the residual nucleus, the energy spectra of the emitted charged-particles are expected to be investigated for the
seventeen photo-induced reactions.

\newcommand{\tabincell}[2]{\begin{tabular}{@{}#1@{}}#2\end{tabular}}
\begin{table*}[htb]
\caption{$S_\mathrm{p}$: the proton separation energies. $S_{\alpha}$: the $\alpha$-particle separation energies.
$E_\mathrm{req.}$: the required energies of the incident $\gamma$-beam for performing the measurements of which the minimum
($\gamma$,p$_\mathrm{tot.}$) or ($\gamma$,$\alpha_\mathrm{tot.}$) yield of 100 counts per day can be attained.
[$E_\mathrm{low}$,$E_\mathrm{upper}$]: the energy range of the incident $\gamma$-beam for performing the measurements, in which the
sum of the yields of 1, 2, and 3 ($\gamma$,$X_{i}$) ($X$=p or $\alpha$) channels are respectively dominant and can be
experimentally determined. Correspondingly, these dominant ($\gamma$,$X_{i}$) channels are explicitly given as well. The unit of all
the energies is MeV. See the text for more details.}

\begin{ruledtabular}
\footnotesize\rm
\begin{spacing}{0.85}
\begin{tabular}{cccccc}

Reaction & $S_\mathrm{p}$ or $S_{\alpha}$ & $E_\mathrm{req}$
& \tabincell{c}{$E_\mathrm{range}$ for $n$ = 1 \\ states}
& \tabincell{c}{$E_\mathrm{range}$ for $n$ = 2 \\ states}
& \tabincell{c}{$E_\mathrm{range}$ for $n$ = 3 \\ states} \\
\colrule
$^{29}$Si($\gamma$,p)$^{28}$Al  & 12.30  & 12.70
&
& \tabincell{c}{ 12.7 $-$ 15.0 \\ p$_{0}$, p$_{1}$}
& \\

$^{50}$V($\gamma$,$\alpha$)$^{46}$Sc & 9.88  &  13.3
&
&
& \tabincell{c}{ 13.3 $-$ 15 \\ $\alpha_{0}$, $\alpha_{1}$, $\alpha_{4}$} \\

$^{56}$Fe($\gamma$,p)$^{55}$Mn   & 10.20 & 10.80
& \tabincell{c}{ 10.8-12 \\ p$_{0}$}
& \tabincell{c}{ 12 $-$ 13.5 \\ p$_{0}$, p$_{1}$}
& \\

$^{74}$Se($\gamma$,p)$^{73}$As &   8.55  & 9.3
&
&
& \tabincell{c}{9.3 $-$ 10.2\\ p$_{0}$, p$_{1}$, p$_{2}$} \\

$^{84}$Sr($\gamma$,p)$^{83}$Rb  &   8.87  & 9.9
&
&
& \tabincell{c}{9.9 $-$ 11.0 \\ p$_{0}$, p$_{1}$, p$_{3}$} \\

$^{87}$Sr($\gamma$,$\alpha$)$^{83}$Kr  & 7.31    & 13.5
&
& \tabincell{c}{13.5 $-$ 14.0 \\ $\alpha_{0}$, $\alpha_{1}$}
& \\

$^{91}$Zr($\gamma$,p)$^{90}$Y     & 8.69     & 10.3
&
& \tabincell{c}{10.3$-$13 \\ p$_{0}$, p$_{1}$}
&  \\

$^{96}$Ru($\gamma$,p)$^{95}$Tc   & 7.35    &8.5
& \tabincell{c}{ 8.5$-$10.5 \\  p$_{1}$}
& \tabincell{c}{ 10.5$-$11.5 \\ p$_{1}$, p$_{4}$}
&  \\

$^{98}$Ru($\gamma$,p)$^{97}$Tc   & 8.29    &9.6
& \tabincell{c}{ 9.6$-$11 \\  p$_{1}$}
& \tabincell{c}{ 11$-$12 \\ p$_{1}$, p$_{4}$}
& \\

$^{102}$Pd($\gamma$,p)$^{101}$Rh  & 7.81    &9.1
& \tabincell{c}{ 9.1$-$10.0 \\ p$_{0}$ }
& \tabincell{c}{ 10.0$-$11.5 \\ p$_{0}$, p$_{3}$}
& \\

$^{106}$Cd($\gamma$,p)$^{105}$Ag     & 7.35    &8.5
& \tabincell{c}{ 8.5$-$10 \\ p$_{0}$}
& \tabincell{c}{ 10$-$11 \\ p$_{0}$, p$_{3}$}
& \\

$^{115}$Sn($\gamma$,p)$^{114}$In     & 8.75    &  11
&
& \tabincell{c}{ 11$-$14 \\ p$_{0}$, p$_{3}$}
& \\

$^{117}$Sn($\gamma$,p)$^{116}$In     & 9.44    &  11.9
&
& \tabincell{c}{ 11.9$-$14.5  \\ p$_{0}$, p$_{3}$}
&  \\

$^{119}$Sn($\gamma$,p)$^{118}$In     & 10.10     &  13.1
&
& \tabincell{c}{ 13.1$-$14.5 \\ p$_{0}$, p$_{2}$}
&  \\

$^{123}$Te($\gamma$,$\alpha$)$^{119}$Sn& 1.53    & 10.5
&
& \tabincell{c}{ 10.5$-$12.5 \\ $\alpha_{0}$, $\alpha_{1}$}
& \\

$^{125}$Te($\gamma$,$\alpha$)$^{121}$Sn& 2.25    & 11.1
&
& \tabincell{c}{ 11.1$-$13 \\ $\alpha_{0}$, $\alpha_{2}$}
& \\

$^{149}$Sm($\gamma$,$\alpha$)$^{145}$Nd &   0.00 &    8.5
&
&
& \tabincell{c}{8.5-10.5 \\ $\alpha_{0}$, $\alpha_{1}$, $\alpha_{2}$}  \\

\label{t1}
\end{tabular}
\end{spacing}
\end{ruledtabular}
\end{table*}

\subsection{Energy spectra}

The peak energy of the emitted particle from the photon-induced reaction $A$($\gamma$,$X_{i}$)$B$ is calculated by
\begin{eqnarray}
E_X = \frac{M_B}{M_B+M_X}(E_{\gamma}+Q-E^{*}_{i}),
\label{simu}
\end{eqnarray}
in which $M_{B}$ and $M_{X}$ are the atomic mass of the residual nucleus and the emitted particle, respectively. In the present case $X$ denotes proton or $\alpha$ particle; $E_{\gamma}$ is the energy of the incident $\gamma$-beam; $Q$ is the Q-value of the reaction $A$($\gamma$,$X$)$B$; and $E^{*}_{i}$ is the excitation energy of the $i$-th state in the residual nucleus $B$. For the photon-induced reaction $A$+$\gamma$ with the emissions of $\alpha$ and proton, the energy gap $\Delta E$ between the peaks of the emitted $\alpha$ and the emitted proton can be expressed as
\begin{eqnarray}
\Delta E &=& E_{\alpha}- E_{p}  \nonumber\\
&=& \frac{M_{(A-\alpha)}}{M_{(A-\alpha)}+M_{\alpha}}(E_{\gamma}+Q_{(\gamma,\alpha)}-E^{*}_{i\ (A-\alpha)})\nonumber\\
&&- \frac{M_{(A-p)}}{M_{(A-p)}+M_{p}}(E_{\gamma}+Q_{(\gamma,p)}-E^{*}_{i\ (A-p)}).
\label{simu1}
\end{eqnarray}

The existence of the energy gap $\Delta E$ is beneficial to the disentanglement of the emitted proton and $\alpha$-particle from a photon-induced reaction, and the larger $\Delta E$ indicates the better particle identification. In Eq. \ref{simu}, the excitation energy $E^{*}$ of the low-lying state is usually much less than the Q-value, so the energy gap $\Delta E$ between the $\alpha$-particle peak and the proton peak is determined by the discrepancy between $Q_{(\gamma,p)}$ and $Q_{(\gamma,\alpha)}$. In this case, it is demonstrated by our previous study \cite{lan2018determination} that, for most photon-induced reactions on the p-nuclei, the emitted proton and $\alpha$-particle can be effectively disentangled from the energy spectra.

Furthermore, according to Eq. \ref{simu}, the energy gap $\Delta E$ between the peaks of two identical charged-particles $X$ ($X$ = proton or $\alpha$-particle), which are respectively emitted from the ($\gamma$,$X_{i}$) channels with two different $i$-th final states in the residual nucleus, is the difference of the excitation energies of these two final states. In the present study, for the seventeen photo-induced reactions listed in Table \ref{t1}, the energy spectra of ($\gamma$,$X_{i}$) are simulated with the energies of the incident $\gamma$-beam in [$E_\mathrm{low}$,$E_\mathrm{upper}$]. This allows us to investigate the possibility of disentangling the identical charged-particles and obtaining the yields of the individual ($\gamma$,$X_{i}$) channels. Note that the minimum measurable limit of 100 counts per day is still taken into account for the total yields obtained from integrating the energy spectrum of each ($\gamma$,$X_{i}$).

\begin{figure*}
\hspace*{-0.2cm}
\centering
\vspace*{-0cm}
\includegraphics[width=12cm,clip]{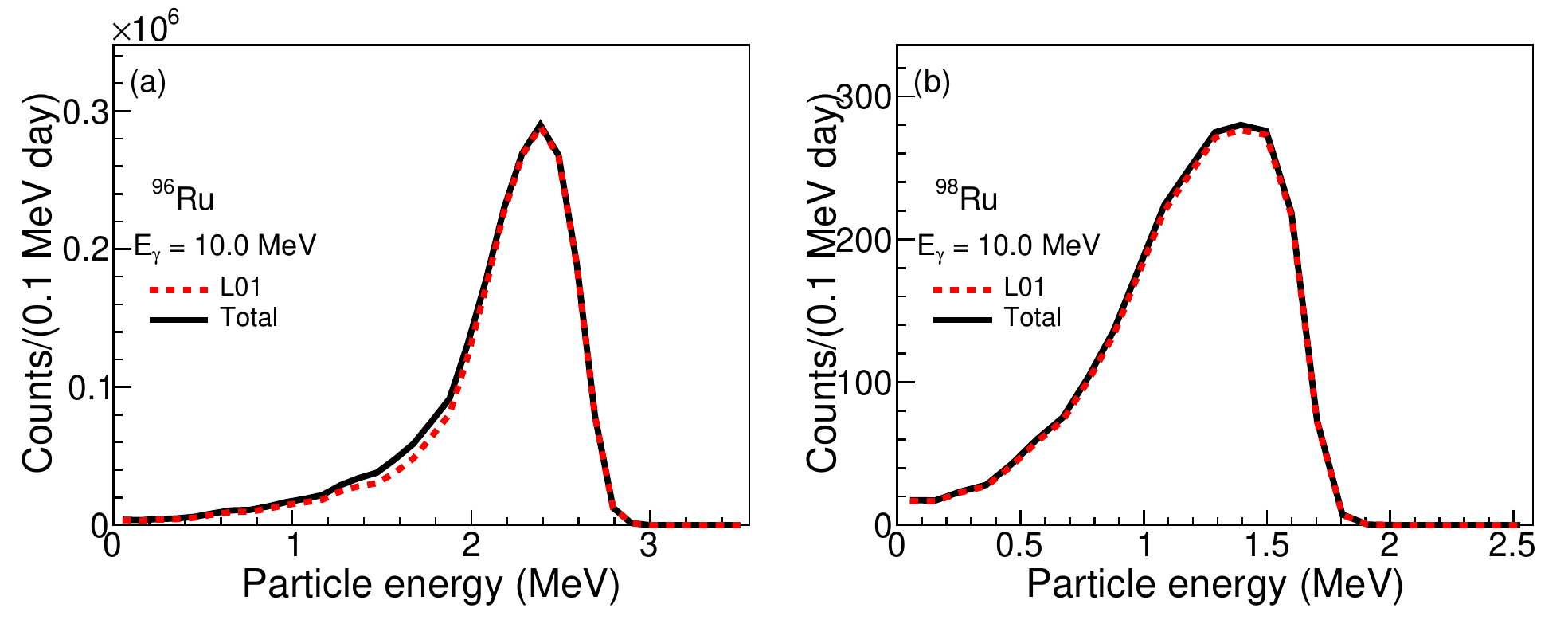}
\caption{\label{f0} Energy spectra of the protons emitted from ($\gamma$,p$_{1}$) and ($\gamma$,p$_\mathrm{tot.}$) channels, indexed
by "L01" and "Total" respectively, on $^{96}$Ru target (a) and $^{98}$Ru target (b) at $E_{\gamma}$ = 10.0 MeV. The results of
($\gamma$,p$_\mathrm{tot.}$) include the contributions from all the ($\gamma$,p$_{i}$) channels with $0\leq i\leq 10$.}
\end{figure*}

For $^{96}$Ru and $^{98}$Ru, the energy spectra of the protons emitted from ($\gamma$,p$_{1}$) and ($\gamma$,p$_\mathrm{tot.}$)
channels at $E_{\gamma}$ = 10.0 MeV are shown in Figure~\ref{f0}. Here ($\gamma$,p$_\mathrm{tot.}$) denotes the proton emission
proceeding on all the eleven final states in the residual nucleus. The results indicate that the protons generated by
$^{96,98}$Ru($\gamma$,p$_{1}$) are dominant at $E_{\gamma}$ $\simeq$ 10.0 MeV ($\approx$ $Q_{(\gamma,p)}$ + 2.5 MeV). The first
excited states in the residual nuclei $^{95,97}$Tc are metastable (isomer) states with the half-lives of 61 and 91 days
respectively, which lead to the relative larger yields and cross sections for ($\gamma$,p$_{1}$) reactions. Thus the measurements
of $^{96,98}$Ru($\gamma$,p$_{1}$) at $E_{\gamma}$ = 10.0 MeV are feasible.

\begin{figure*}
\hspace*{-0.2cm}
\centering
\vspace*{-0cm}
\includegraphics[width=12cm,clip]{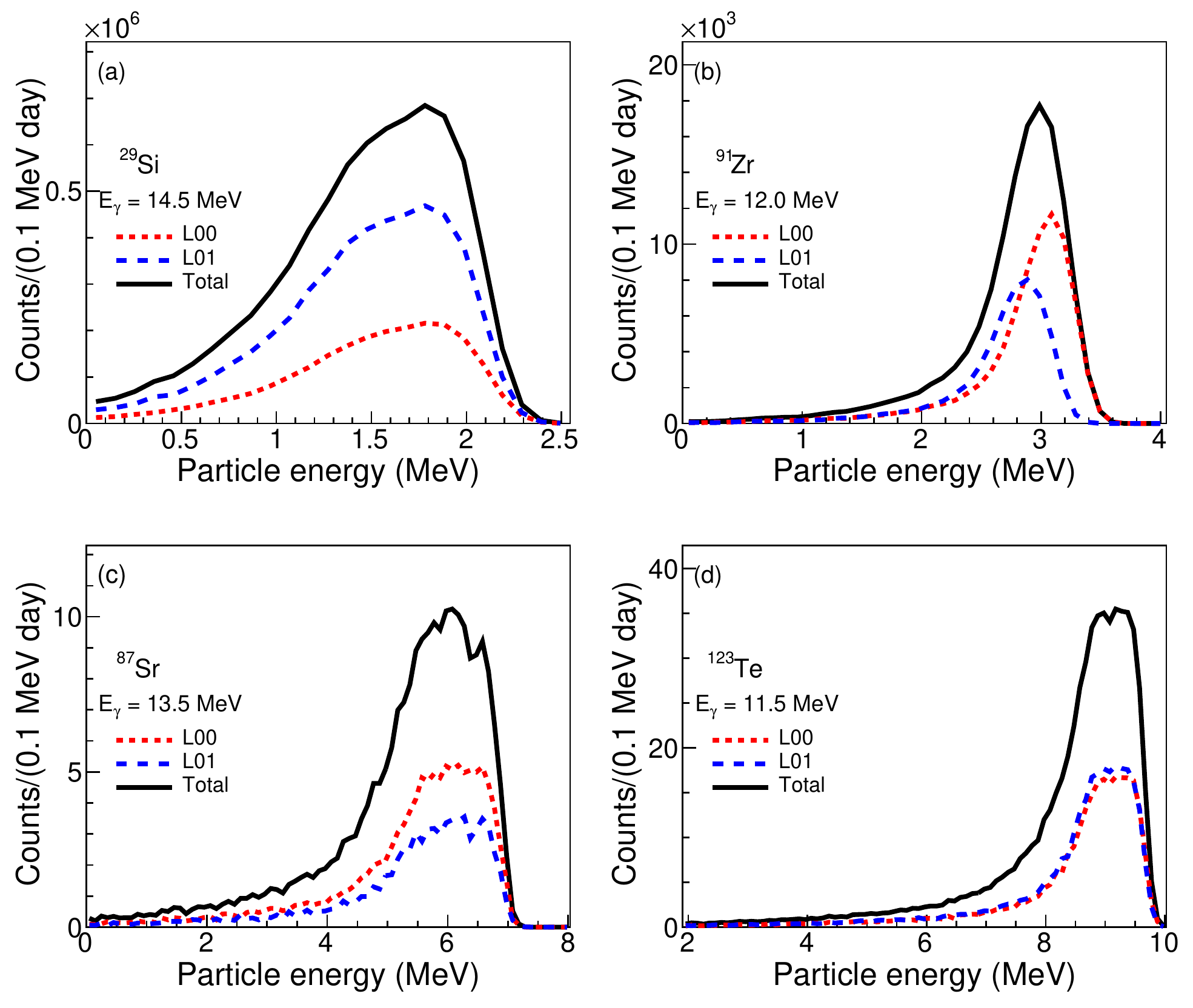}
\caption{\label{f2} Energy spectra of the protons emitted from ($\gamma$,p$_{0}$), ($\gamma$,p$_{1}$) and
($\gamma$,p$_\mathrm{tot.}$) channels on $^{29}$Si at $E_{\gamma}$ = 14.5 MeV (a) and $^{91}$Zr at $E_{\gamma}$ = 12.0 MeV (b), and
energy spectra of the $\alpha$-particles emitted from ($\gamma$,$\alpha_{0}$), ($\gamma$,$\alpha_{1}$) and
($\gamma$,$\alpha_\mathrm{tot.}$) channels on $^{87}$Sr at $E_{\gamma}$ = 13.5 MeV (c) and $^{123}$Te at $E_{\gamma}$ = 11.5 MeV
(d). The results of ($\gamma$,$X_\mathrm{tot.}$) ($X$ = p or $\alpha$) include the contributions from all the ($\gamma$,$X_{i}$)
channels with $0\leq i\leq 10$.}
\end{figure*}

Figure~\ref{f2} demonstrates the proton spectra of ($\gamma$,p$_{0}$), ($\gamma$,p$_{1}$) and ($\gamma$,p$_\mathrm{tot.}$) on
$^{29}$Si target at $E_{\gamma}$ = 14.5 MeV (a) and $^{91}$Zr target at $E_{\gamma}$ = 12.0 MeV (b), and the $\alpha$-particle
spectra of ($\gamma$,$\alpha_{0}$), ($\gamma$,$\alpha_{1}$) and ($\gamma$,$\alpha_\mathrm{tot.}$) on $^{87}$Sr target at
$E_{\gamma}$ = 13.5 MeV (c) and $^{123}$Te target at $E_{\gamma}$ = 11.5 MeV (d). From these four reactions, the reaction channels
leading to the ground state and the first excited state in the residual nucleus contribute more than 90$\%$ to the total spectra of
the emitted charged-particles. It can be also found that the sum of the yields of ($\gamma$,$X_{0}$) and ($\gamma$,$X_{1}$) is
measurable, while the yields of the two respective channels cannot be experimentally determined.

\begin{figure*}[!htb]
\centering
\includegraphics[width=12cm,clip]{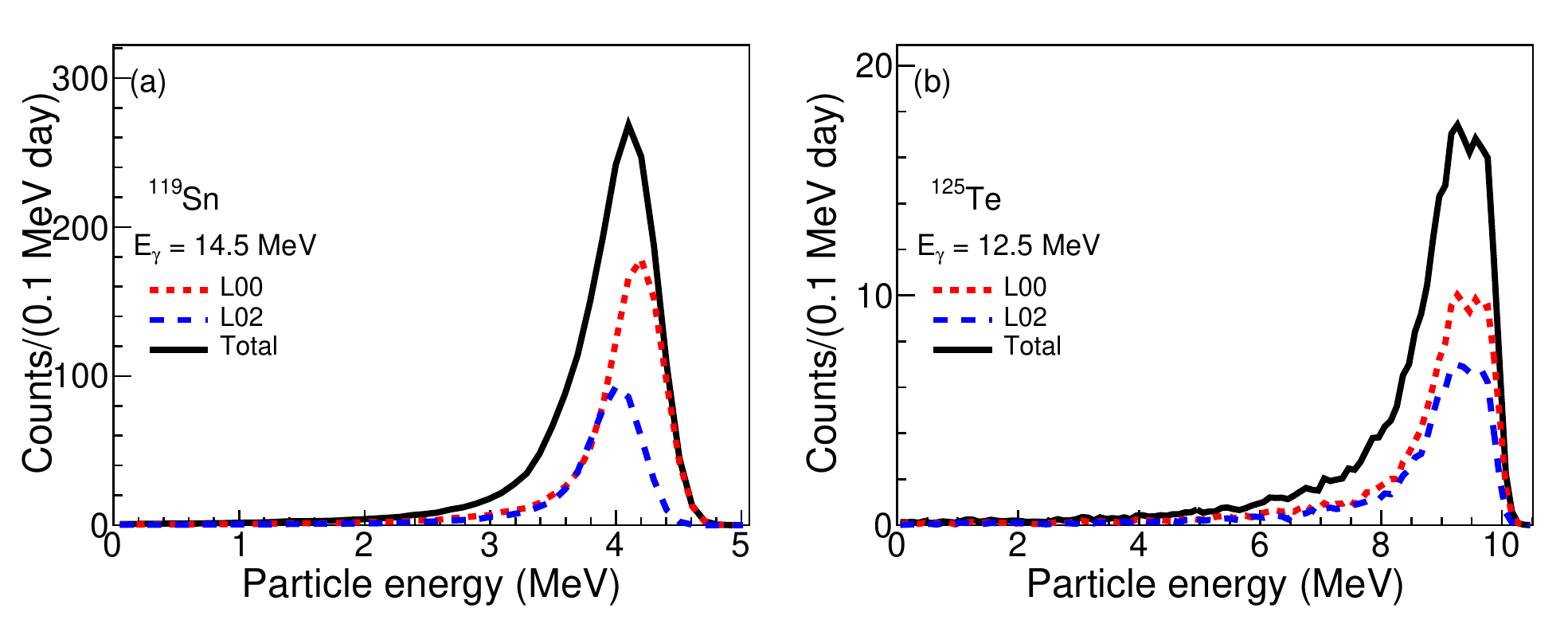}
\caption{\label{fl02} Energy spectra of the protons emitted from ($\gamma$,p$_{0}$), ($\gamma$,p$_{2}$) and
($\gamma$,p$_\mathrm{tot.}$) channels on $^{119}$Sn at $E_{\gamma}$ = 14.5 MeV (a), and energy spectra of the $\alpha$-particles
emitted from ($\gamma$,$\alpha_{0}$), ($\gamma$,$\alpha_{2}$) and ($\gamma$,$\alpha_\mathrm{tot.}$) channels on $^{125}$Te at
$E_{\gamma}$ = 12.5 MeV (b). The results of ($\gamma$,$X_\mathrm{tot.}$) ($X$ = p or $\alpha$) include the contributions from all
the ($\gamma$,$X_{i}$) channels with $0\leq i\leq 10$.}
\end{figure*}

In Figure~\ref{fl02} the proton spectra of ($\gamma$,p$_{0}$), ($\gamma$,p$_{2}$) and ($\gamma$,p$_\mathrm{tot.}$) on $^{119}$Sn at
$E_{\gamma}$ = 14.5 MeV (a) and the $\alpha$-particle spectra of ($\gamma$,$\alpha_{0}$), ($\gamma$,$\alpha_{12}$) and
($\gamma$,$\alpha_\mathrm{tot.}$) on $^{125}$Te at $E_{\gamma}$ = 12.5 MeV (b) are given. It is shown that the sum of the yields of
($\gamma$,$X_{0}$) and ($\gamma$,$X_{2}$) can be measured because of their dominant contributions to the total yields of all the
eleven transitions. However, the contributions of ($\gamma$,$X_{0}$) and ($\gamma$,$X_{2}$) are not
experimentally distinguishable. According to the simulation, $^{125}$Te($\gamma$,$\alpha_{2}$) contributes 40$\%$ to the total
$\alpha$-particle yield at 11.0 $<$ $E_{\gamma}$ $<$ 12.5 MeV, in which the second excited state of the residual nucleus $^{121}$Sn
is an isomeric state with a long half-life of 43.9 years.

\begin{figure*}[!htb]
\centering
\includegraphics[width=12cm,clip]{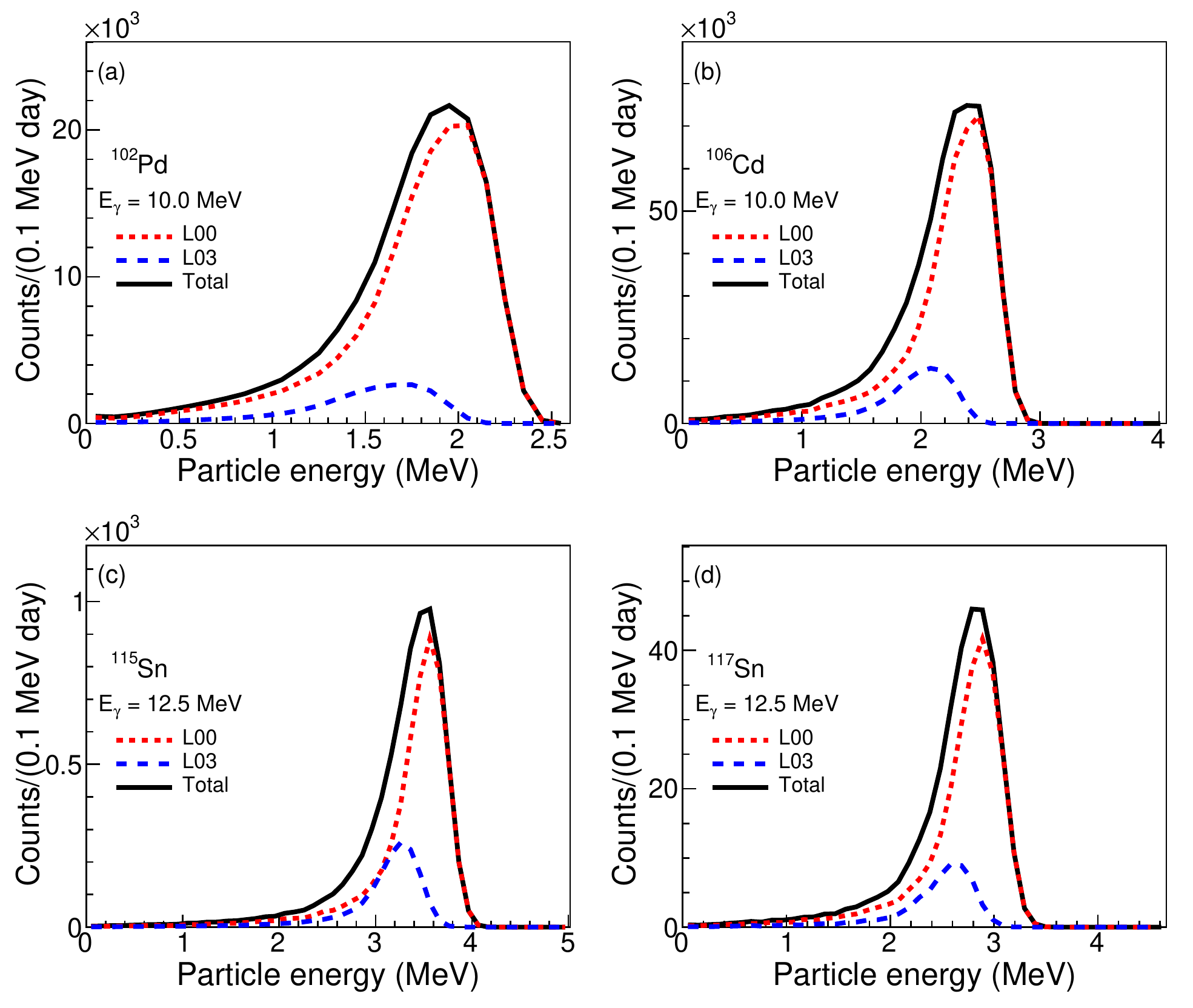}
\caption{\label{fl03} Energy spectra of the protons emitted from ($\gamma$,p$_{0}$), ($\gamma$,p$_{3}$) and
($\gamma$,p$_\mathrm{tot.}$) channels on $^{102}$Pd (a) and $^{106}$Cd (b) at $E_{\gamma}$ = 10.0 MeV, and $^{115}$Sn (c) and
$^{117}$Sn (d) at $E_{\gamma}$ = 12.5 MeV. The results of ($\gamma$,p$_\mathrm{tot.}$) include the contributions from all the
($\gamma$,$X_{i}$) channels with $0\leq i\leq 10$.}
\end{figure*}

Figure~\ref{fl03} illustrates the proton spectra of ($\gamma$,p$_{0}$), ($\gamma$,p$_{3}$) and ($\gamma$,p$_\mathrm{tot.}$) on
$^{102}$Pd (a) and $^{106}$Cd (b) at $E_{\gamma}$ = 10.0 MeV, and $^{115}$Sn (c) and $^{117}$Sn (d) at $E_{\gamma}$ = 12.5 MeV. For
these four targets, the sum of the yields of ($\gamma$,p$_{0}$) and ($\gamma$,p$_{3}$) can be measured, which prominently
contributes the total proton yields. Still, the energy spectra show that the respective contributions of ($\gamma$,p$_{0}$) and
($\gamma$,p$_{3}$) cannot be recognized.

\begin{figure*}
\centering
\includegraphics[width=12cm,clip]{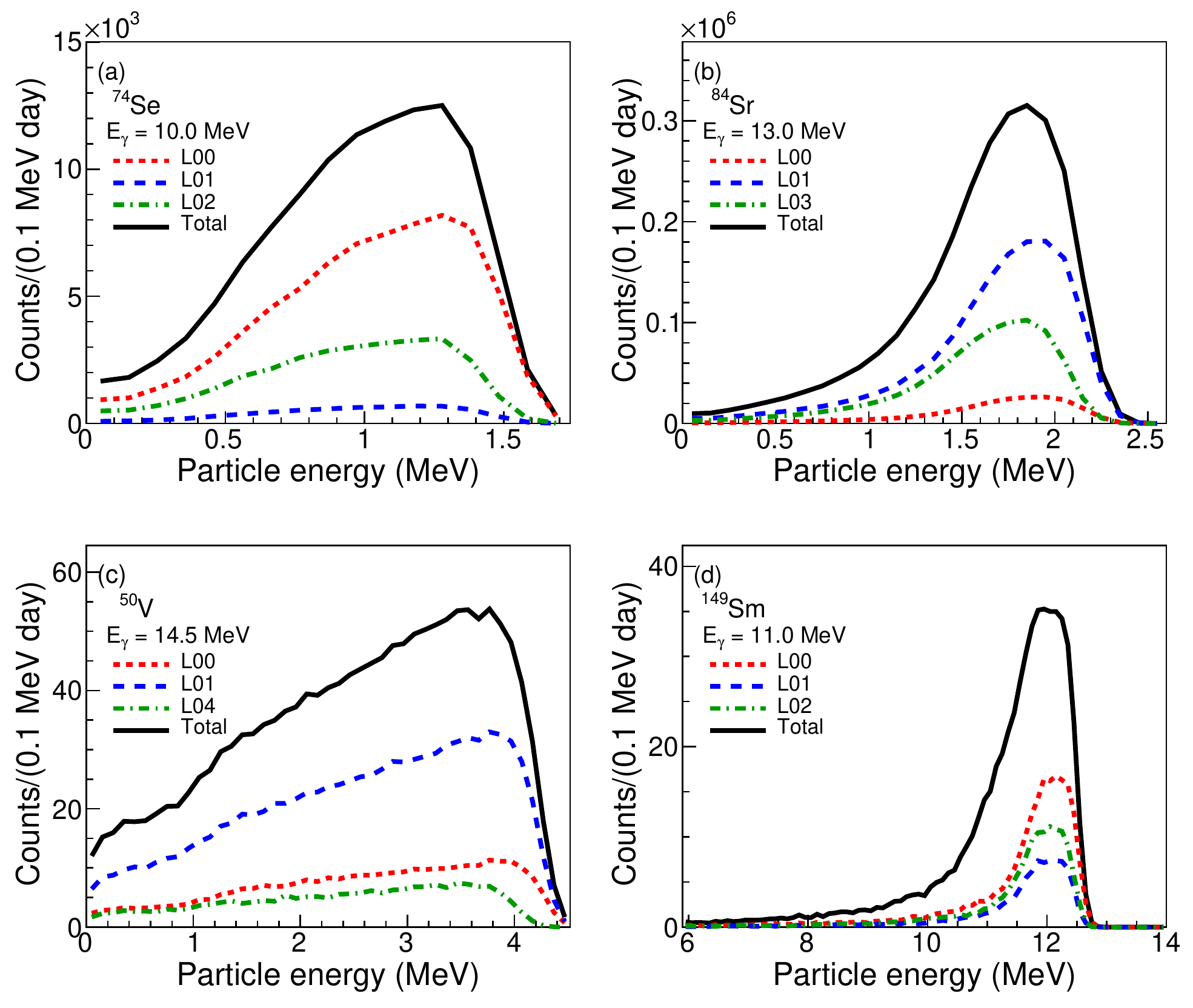}
\caption{\label{f3e} Energy spectra of the protons emitted from ($\gamma$,p$_{0}$), ($\gamma$,p$_{1}$), ($\gamma$,p$_{2}$) and
($\gamma$,p$_\mathrm{tot.}$) channels on $^{74}$Se at $E_{\gamma}$ = 10.0 MeV (a), and from ($\gamma$,p$_{0}$), ($\gamma$,p$_{1}$),
($\gamma$,p$_{3}$) and ($\gamma$,p$_\mathrm{tot.}$) on $^{84}$Sr at $E_{\gamma}$ = 13.0 MeV (b). Energy spectra of the
$\alpha$-particles emitted from ($\gamma$,$\alpha_{0}$), ($\gamma$,$\alpha_{1}$), ($\gamma$,$\alpha_{4}$) and
($\gamma$,$\alpha_\mathrm{tot.}$) on $^{50}$V at $E_{\gamma}$ = 14.5 MeV (c), and ($\gamma$,$\alpha_{0}$), ($\gamma$,$\alpha_{1}$),
($\gamma$,$\alpha_{2}$) and ($\gamma$,$\alpha_\mathrm{tot.}$) on $^{149}$Sm+$\gamma$ at $E_{\gamma}$ = 11.0 MeV (d). The results of
($\gamma$,$X_\mathrm{tot.}$) (X = p or $\alpha$) include the contributions from all the ($\gamma$,$X_{i}$) channels with
$0\leq i\leq 10$.}
\end{figure*}

In Figure~\ref{f3e} the proton spectra of ($\gamma$,p$_{0}$), ($\gamma$,p$_{1}$), ($\gamma$,p$_{2}$) and ($\gamma$,p$_\mathrm{tot.}$)
on $^{74}$Se at $E_{\gamma}$ = 10.0 MeV (a), and ($\gamma$,p$_{0}$), ($\gamma$,p$_{1}$), ($\gamma$,p$_{3}$) and
($\gamma$,p$_\mathrm{tot.}$) on $^{84}$Sr at $E_{\gamma}$ = 13.0 MeV (b), as well as the $\alpha$-particle spectra of
($\gamma$,$\alpha_{0}$), ($\gamma$,$\alpha_{1}$), ($\gamma$,$\alpha_{4}$) and ($\gamma$,$\alpha_\mathrm{tot.}$) on $^{50}$V at
$E_{\gamma}$ = 14.5 MeV (c), and ($\gamma$,$\alpha_{0}$), ($\gamma$,$\alpha_{1}$), ($\gamma$,$\alpha_{2}$) and
($\gamma$,$\alpha_\mathrm{tot.}$) on $^{149}$Sm+$\gamma$ at $E_{\gamma}$ = 11.0 MeV (d) are illustrated. For these four reactions,
the total yields are contributed dominantly from three individual channels with the transitions to the ground and two excited states
in the residual nucleus, so the measured yields are the sum of the yields of these three channels. However, each of the contributions from the three transitions cannot be respectively identified from the measurement.

\begin{figure*}
\centering
\includegraphics[width=12cm,clip]{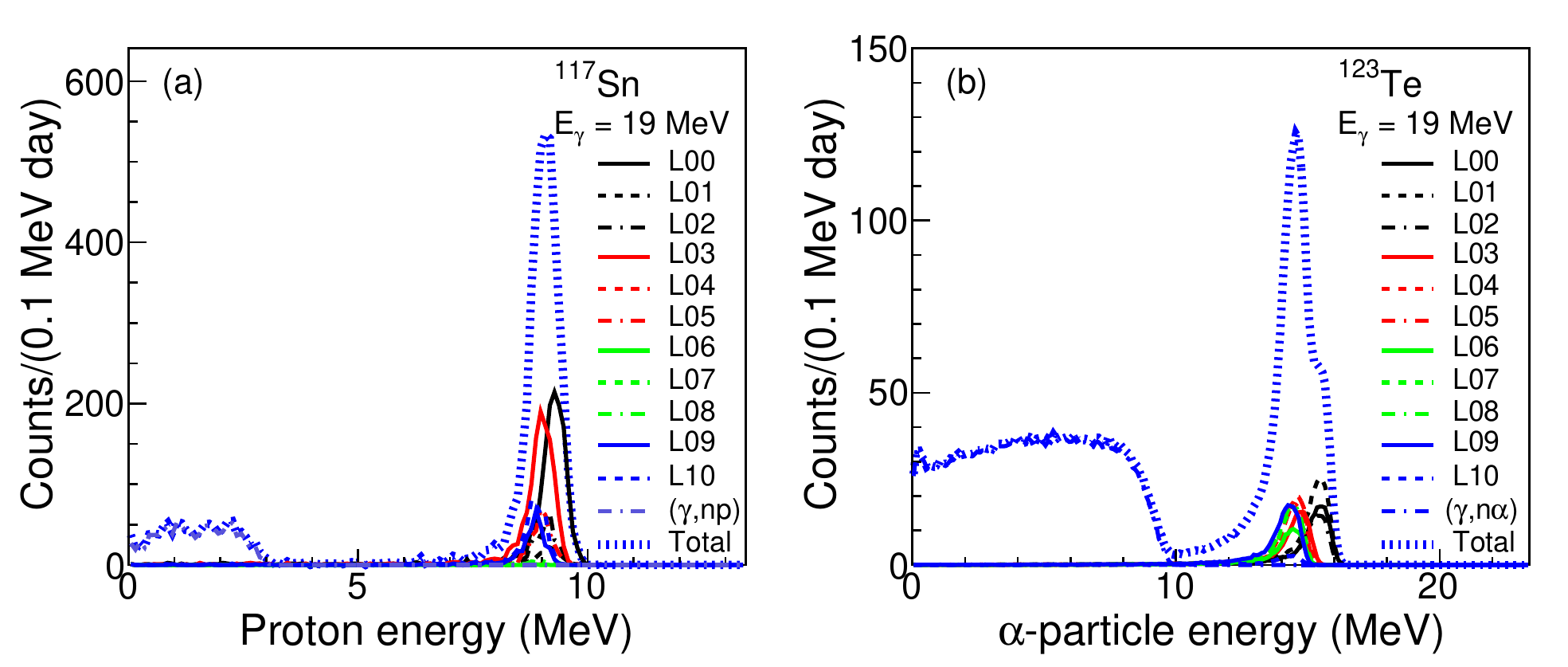}
\caption{\label{fmpe} The energy spectra of the proton from $^{117}Sn$+$\gamma$ reaction and the $\alpha$-particle from
$^{123}Te$+$\gamma$ reaction at $E_{\gamma}$ = 19.0 MeV. The results of ($\gamma$,$X_{i}$) channels are indexed by "L0i" for
$0\leq i\leq 10$, and the results of ($\gamma$,$np$) and ($\gamma$,$n$$\alpha$) channels are indexed by "($\gamma$,$np$)" and
"($\gamma$,$n$$\alpha$)", respectively. The label "Total" denotes the results counting all these considered contributions.}
\end{figure*}

Figure~\ref{fmpe} demonstrates the proton spectra from $^{117}Sn$+$\gamma$ reaction and the $\alpha$-particle spectra from
$^{123}Te$+$\gamma$ reaction at $E_{\gamma}$ = 19.0 MeV. The multiple emission channels ($\gamma$,$np$) and ($\gamma$,$n$$\alpha$)
are taken into account. It is found the energies of the emitted charged particles from the multiple channels are remarkably lower
than the energies of the particles from ($\gamma$,$X_{i}$). Therefore, although the cross sections of multiple emissions are
comparable to those of ($\gamma$,$X_{i}$), the multiple emission events can be readily removed by setting a proper threshold for the
detection of the particle energies, and the interesting measurements of ($\gamma$,$X_{i}$) cannot be influenced.

\section{Summary}
\label{sec:summary}

Capture and photodisintegration reactions proceeding on excited states of the target nucleus can significantly contribute to the astrophysical reaction rate. Those excited-state contributions are usually obtained from theoretical calculations but so far are inaccessible to a direct measurement.
It is, however, possible to test and constrain theoretical models by measuring relevant nuclear properties in dedicated experiments, which can then provide better predictions of reaction rates. In this paper, we propose to measure the photon-induced reactions with proton and $\alpha$-particle emissions, to investigate the nuclear transitions leading to specific excited states in the residual nucleus. The particle-transmission coefficients of the charged-particles for photodisintegration and capture reactions can be determined by the anticipated experimental result, which is important to improve the predictions of astrophysical reaction rates for astrophysical simulation of $p$ nuclides production.
For the ($\gamma$,p$_{i}$) reactions on seventeen targets and the $(\gamma,\alpha_{i})$ reactions on seventeen targets, the sensitivity of the calculated cross sections to the nuclear properties is studied. It is found that the $\alpha$-particle optical model potentials can considerably influence the relative contributions of ($\gamma$,$\alpha_{i}$) cross sections to the total cross sections of ($\gamma$,$\alpha_\mathrm{tot.}$).

At the ELI-NP facility, a Variable Energy $\gamma$-ray (VEGA) system is being constructed, which opens
new opportunities to experimentally study the photon-induced reactions of astrophysics interest. Based on the features of VEGA and the silicon strip array ELISSA developed at ELI-NP for charged-particle detection, GEANT4 simulations of the measurements of the photon-induced reactions with the proton emission on seventeen targets and the $\alpha$-particle emission on seventeen targets are performed using the calculated cross sections. The yields of ($\gamma$,p$_{i}$) and ($\gamma$,$\alpha_{i}$) channels are obtained from the simulation. The results reveal that, for the ($\gamma$,p) reaction on twelve targets ($^{29}$Si, $^{56}$Fe, $^{74}$Se, $^{84}$Sr, $^{91}$Zr, $^{96,98}$Ru, $^{102}$Pd, $^{106}$Cd, and $^{115, 117, 119}$Sn) and the ($\gamma$,$\alpha$) reaction on five targets ($^{50}$V, $^{87}$Sr, $^{123,125}$Te, and $^{149}$Sm), the yields of ($\gamma$,$X_{i}$) ($X$=p or $\alpha$) channels with $i\neq0$, namely the transitions to the exited states in the residual nucleus, are relevant and even dominant. Therefore, the feasibility of measuring these seventeen photo-induced reactions at VEGA+ELISSA is investigated further.

For each of the seventeen photon-induced reactions, in order to attain the minimum detectable limit of 100 counts per day for the total proton or $\alpha$-particle yields, the minimum required energies of the incident $\gamma$-beam ($E_\mathrm{low}$) for the measurements are estimated. It is further found that for each considered reaction, the total yield of the charged-particle $X$ may be dominantly contributed from one, two or three individual ($\gamma$,$X_{i}$) channels within a specific energy range of the incident $\gamma$-beam, i.e., [$E_\mathrm{low}$,$E_\mathrm{upper}$]. If each of these photon-induced reactions are actually measured within their respective energy ranges, the sum of the yields of the dominant ($\gamma$,$X_{i}$) channels can be approximated by the measured yields of the charged particle $X$ within acceptable uncertainty. This means that the yields of the ($\gamma$,$X_{i}$) channels prominently contributing to the total yields of $X$ can be experimentally obtained. Therefore, it is important to determine the interval [$E_\mathrm{low}$,$E_\mathrm{upper}$]. Using the simulated yields, such energy ranges [$E_\mathrm{low}$,$E_\mathrm{upper}$] for each of the seventeen photon-induced reactions are derived.

Furthermore, for these seventeen targets, the energy spectra of ($\gamma$,$X_{i}$) channels are simulated with the incident $\gamma$-beam energies in their interval [$E_\mathrm{low}$,$E_\mathrm{upper}$]. Based on the energy
spectra, the identifications of the individual dominant ($\gamma$,$X_{i}$) channels are discussed. It is shown that the
($\gamma$,p$_{0}$) reaction channel for the targets of $^{56}$Fe, $^{102}$Pd and $^{106}$Cd, and the ($\gamma$,p$_{1}$) channel for the targets of $^{96}$Ru and $^{98}$Ru can be exclusively determined. Therefore it becomes evident that a measurement of the photon-induced reactions with charged-particle emissions considered in this work is feasible with the VEGA+ELISSA system and will provide knowledge useful for nuclear astrophysics.

\begin{acknowledgments}
This work is supported by the National Natural Science Foundation of China (Grant No. 11675075). This work is carried out under the
contract PN 19 06 01 05 sponsored by the Romanian Ministry of Research and Innovation. Y.X., D.L.B. and C.M. acknowledge supports from the
ELI-RO program funded by the Institute of Atomic Physics (Magurele, Romania) under the contract ELI\_15/16.10.2020, and from the
Extreme Light Infrastructure Nuclear Physics (ELI-NP) - Phase II, a project co-financed by the Romanian Government and the European
Union through the European Regional Development Fund - the Competitiveness Operational Programme (1/07.07.2016, COP, ID 1334). T.R.
is partially supported by the European COST action "ChETEC" (CA16117). This work was financially supported by the Italian Ministry of Education, University and Research, PON R\&I 2014-2020 - AIM, project AIM1848704-3.
\end{acknowledgments}

\section*{Reference}
\bibliography{ref}
\end{document}